\PassOptionsToPackage{table,dvipsnames}{xcolor}
\PassOptionsToPackage{hyphens}{url}
\documentclass[lettersize,journal]{IEEEtran}
\IEEEoverridecommandlockouts

\usepackage{soul}

\newcommand{\revise}[1]{{\color{blue} #1}}
\renewcommand{\revise}[1]{#1}

\usepackage[left=0.625in,right=0.625in,top=0.705in,bottom=1.02in]{geometry}
\usepackage{cite}
\usepackage{amsmath,amssymb,amsfonts,amsthm}
\usepackage{algorithmic}
\usepackage{graphicx}
\usepackage{textcomp}
\usepackage{xcolor}
\usepackage{enumitem}
\usepackage{mathtools}
\usepackage{comment}
\PassOptionsToPackage{hyphens}{url}\usepackage[hidelinks]{hyperref}
\usepackage{tikz,pgfplots}
\usepackage{grffile}
\usepgfplotslibrary{patchplots}
\usetikzlibrary{patterns,backgrounds,plotmarks,fit,positioning,arrows,shapes,shapes.multipart,calc,arrows.meta}
\usetikzlibrary{fit,positioning,arrows,shapes,shapes.multipart,calc,arrows.meta,shapes.geometric,shapes.misc}
\usetikzlibrary{decorations.pathreplacing,angles,quotes,calligraphy}
\usetikzlibrary{decorations.markings}
\usetikzlibrary{automata}

\allowdisplaybreaks

\usepackage[labelformat=simple]{subcaption}

\usepackage[toc,acronym,nonumberlist,nopostdot]{glossaries}
\makeatletter
\patchcmd{\@gls@}
{\@gls@link[#1]{#2}{\@glo@text}}
{\@gls@link[#1,hyper=false]{#2}{\@glo@text}}
{}{}

\patchcmd{\@glspl@}
{\@gls@link[#1]{#2}{\@glo@text}}
{\@gls@link[#1,hyper=false]{#2}{\@glo@text}}
{}{}

\patchcmd{\@Gls@}
{\@gls@link[#1]{#2}}
{\@gls@link[#1,hyper=false]{#2}}
{}{}

\patchcmd{\@GLS@}
{\@gls@link[#1]{#2}{\MakeUppercase{\@glo@text}}}
{\@gls@link[#1,hyper=false]{#2}{\MakeUppercase{\@glo@text}}}
{}{}
\makeatother

\newacronym{IoT}{IoT}{Internet of Things}
\newacronym{MAC}{MAC}{multiple access channel}
\newacronym{AWGN}{AWGN}{additive white Gaussian noise}
\newacronym{SNR}{SNR}{signal-to-noise ratio}
\newacronym{SINR}{SINR}{signal-to-interference-plus-noise ratio}
\newacronym{AP}{AP}{access point}
\newacronym{UE}{UE}{user equipment}
\newacronym{CPU}{CPU}{central processing unit}
\newacronym{OFDM}{OFDM}{orthogonal frequency division multiplexing}
\newacronym{TDD}{TDD}{time-division duplexing}
\newacronym{RF}{RF}{radio frequency}
\newacronym{LO}{LO}{local oscillator}
\newacronym{IID}{IID}{independent and identically distributed}
\newacronym{MMSE}{MMSE}{minimum mean squared error}
\newacronym{LMMSE}{LMMSE}{linear minimum mean squared error}
\newacronym{MSE}{MSE}{mean squared error}
\newacronym{CP}{CP}{cyclic prefix}
\newacronym{DFT}{DFT}{discrete Fourier transform}
\newacronym{IDFT}{IDFT}{inverse discrete Fourier transform}
\newacronym{PN}{PN}{phase noise}
\newacronym{CPE}{CPE}{common phase error}
\newacronym{ICI}{ICI}{inter-carrier interference}
\newacronym{ISI}{ISI}{inter-symbol interference}
\newacronym{IUI}{IUI}{inter-user interference}
\newacronym{SE}{SE}{spectral efficiency}
\newacronym{UatF}{UatF}{use-and-then-forget}
\newacronym{UL}{UL}{uplink}
\newacronym{DL}{DL}{downlink}
\newacronym{iid}{i.i.d.}{independent and identically distributed}

\newacronym{PDF}{PDF}{probability density function}
\newacronym{CDF}{CDF}{cumulative distribution function}
\newacronym{MR}{MR}{maximum ratio}
\newacronym{ZF}{ZF}{zero-forcing}
\newacronym{MIMO}{MIMO}{multiple-input multiple-output}

\usepackage{vmr-symbols-vecbold-RanDetSame}
\usepackage{standard-macros}

\newcommand{\tc}{\tau\sub{c}}
\newcommand{\tpl}{\tau\sub{p}}
\newcommand{\tu}{\tau\sub{u}}
\newcommand{\td}{\tau\sub{d}}

\newcommand{\ts}{\tau\sub{s}}
\newcommand{\tg}{\tau\sub{g}}

\newcommand{\nur}{\nu^{(\rm r)}}
\newcommand{\nut}{\nu^{(\rm t)}}
\newcommand{\deltar}{\delta^{(\rm r)}}
\newcommand{\deltat}{\delta^{(\rm t)}}

\newcommand{\DS}{{\sf DS}}
\newcommand{\BU}{{\sf BU}}
\newcommand{\UI}{{\sf UI}}

\newcommand{\nsync}{F} 
\newcommand{\nmeas}{n\sub{m}}

\newcommand{\SEdl}{{\sf SE}}

\newcommand{\Pdl}{\rho\sub{AP}}
\newcommand{\Ppl}{\rho\sub{UE}}

\def\BibTeX{{\rm B\kern-.05em{\sc i\kern-.025em b}\kern-.08em
    T\kern-.1667em\lower.7ex\hbox{E}\kern-.125emX}}
    
\begin{document}

\title{Distributed MIMO With Over-the-Air Phase Calibration Integrated Into the TDD Flow}

\author{Khac-Hoang Ngo, \emph{Member, IEEE}, and Erik G. Larsson, \emph{Fellow, IEEE} 
	\thanks{{The authors are with the Department of Electrical Engineering (ISY), Linköping University, 58183 Linköping, Sweden~(email: \{khac-hoang.ngo, erik.g.larsson\}@liu.se}).}
	\thanks{This paper was presented in part at the 2025 IEEE Global Communications Conference (GLOBECOM)~\cite{Ngo25breakingTDD}.
    }
    \thanks{This work was supported by the
Excellence Center at Linköping–Lund in Information Technology (ELLIIT),  the Swedish Research Council (VR),
    and the Knut and Alice Wallenberg (KAW) Foundation.}
}

\maketitle

\begin{abstract}
Reciprocity-based, joint coherent  downlink beamforming from multiple access points (APs) in  distributed multiple-input multiple-output (MIMO) with independent
local oscillators (LOs) requires the APs to be periodically phase-calibrated (a.k.a.\ phase-synchronized or phase-aligned). Such phase \revise{calibration} can be accomplished by bidirectional over-the-air
measurements between the APs. In this paper, we show how such over-the-air
measurements can be integrated into the time-division duplexing (TDD) flow by
appropriately shifting the uplink/downlink switching points of the TDD slot structure, 
 creating short time segments during which APs can measure on one another.
We also show how this technique scales to large networks.  
Furthermore, we analytically characterize the tradeoff between  the amount of resources spent on calibration measurements
and the resulting spectral efficiency of the system, when conjugate beamforming or zero-forcing beamforming
is used.
The results demonstrate the feasibility of distributed MIMO with phase-calibration through over-the-air inter-AP measurements integrated into the TDD flow, \revise{and the advantage of this design over schemes with dedicated calibration slots}.

\end{abstract}

\section{Introduction} \label{sec:intro}


Distributed antenna architectures~\cite{Ngo17_CF,Demir21} are widely regarded as a key technology component of the physical layer in sixth-generation (6G) wireless networks. In a distributed antenna system, \glspl{AP}\textemdash each with a single or multiple antennas\textemdash are spread out geographically and cooperate to serve \glspl{UE} through phase-coherent transmission.  Time-division duplexing (TDD)
operation allows the system to exploit channel reciprocity and perform downlink beamforming based on uplink channel estimates. However, when the \glspl{AP} are driven by different \glspl{LO} that are not phase-locked, this reciprocity breaks down, and the signals transmitted from different \glspl{AP} fail to combine coherently \cite{Nissel22,Larsson23_calibration}. In this case, phase calibration\footnote{Also called phase synchronization and phase alignment in the literature.} is required for joint 
beamforming across \glspl{AP}. This 
calibration must be repeated frequently enough to compensate for the LO phase drift over time.
One way of achieving calibration is to perform bi-directional over-the-air measurements between access points. 

In \gls{TDD}, dedicated slots can be used only for calibration. An alternative approach that we pursue in this paper is introducing phase calibration measurements within slots where data are also transmitted. This approach allows to place the measurements evenly in time. However, it disrupts the \gls{TDD} flow. Specifically, over a part of a slot assigned for downlink transmission, if an \gls{AP} transmits a phase calibration signal, the other \glspl{AP} must switch from downlink to uplink mode in order to receive this signal. 
Similarly, if the calibration signal is transmitted during a uplink period, the transmitting \gls{AP} must switch to downlink mode. These uplink/downlink mode switches introduce several design challenges. First, the uplink/downlink switch has to be carefully designed to respect the guard intervals between uplink and downlink transmissions, and to minimize disruption to the TDD flow. Second, complete bidirectional phase measurements between all \glspl{AP} may require a long sequence of slots, during which the APs' \glspl{LO} continue to drift. To minimize the number of slots needed, the \glspl{AP} should be carefully scheduled so that multiple APs can transmit calibration signals simultaneously with minimal interference. 
Third, the impact of measurement errors and the LOs' drift on downlink beamforming should be rigorously characterized via a performance metric, in order to analyze the impact of calibration overhead and guide the selection of key system parameters 
and the design of AP scheduling. This paper aims to address these open challenges.

    


\subsection{Over-the-Air Reciprocity Calibration}
We consider a distributed antenna system 
with multiple \glspl{AP}, each driven by a single, independent \gls{LO} that drifts over time, resulting in a time-varying shift between the LO phases at different \glspl{AP}. We assume that the transmit and receive radio-frequency chains are hardware-balanced, i.e., they induce identical amplitude and phase responses, so that any residual phase variation is solely due to the shared oscillator’s phase noise, which affects both uplink and downlink directions equally.
The transmitted and received baseband signals of AP~$\ell$ at time $i$ are multiplied with the complex-valued coefficients \revise{$\exp(-\jmath \nut_{\ell,i})$} and \revise{$\exp(\jmath \nur_{\ell,i})$}, respectively. 
To enable reciprocity-based downlink beamforming to \gls{UE}~$k$ at time $i$ using its uplink channel estimate obtained at time $[i]_k$, the system needs to know \revise{$\vecphi_i = \{\nut_{\ell, i} + \nur_{\ell, [i]_k}\}_\ell \mod 2\pi$,} up to a common additive constant for all~$\ell$. 
Bidirectional measurements between \gls{AP}~$\ell_1$ and \gls{AP}~$\ell_2$ at times $i_1$ and $i_2$ give the value of \revise{$(\nut_{\ell_2, i_1} + \nur_{\ell_2, i_2}) - (\nur_{\ell_1, i_1} + \nut_{\ell_1, i_2}) \mod 2\pi$.} Let $\vecalpha_i$ collect these values 
up to time $i$ for all \glspl{AP} pairs $(\ell_1,\ell_2)$ involved in the measurements. Over-the-air reciprocity-calibration consists in obtaining $\vecalpha_i$ and finding $\vecphi_i$ therefrom, subject to three noise sources: i) measurement noise in $\vecalpha_i$, ii) phase drift during the time offsets between $i$, $[i]_k$, $i_1$, and $i_2$, iii) estimation error when solving for $\vecphi_i$ from $\vecalpha_i$ under circular nonlinearity.

\subsection{\revise{Related Work}}\label{sec:relatedwork}

Reciprocity-based joint coherent transmission from distributed antennas hinges on both i) \emph{intra}-AP calibration that removes the transmit/receive mismatches per \gls{RF} chain \revise{and the mismatches between the antennas within an \gls{AP}}
so uplink channel estimates can be used for downlink beamforming, and ii) \emph{inter}-\gls{AP} calibration that ensures that signals transmitted from geographically separated \glspl{AP} add coherently at the \gls{UE}. 

\revise{\emph{Intra}-AP reciprocity calibration (within an AP) 
can be accomplished 
either by using additional hardware to
measure  the imperfections of each radio-frequency chain~\cite{Bourdoux03,Benzin17},
or by bidirectional over-the-air signaling between the different antennas of an AP~\cite{Kaltenberger10}.
Experimental investigations of over-the-air calibration for co-located arrays have been reported in~\cite{Zetterberg2010experimental,Jiang15}. The calibration scheme proposed in~\cite{vieira2017reciprocity} exploits mutual coupling between the antennas, while~\cite{Jiang18} further explores antenna grouping to improve accuracy and speed up the calibration process. The authors of~\cite{Luo19} derive the
calibration performance, in terms of the Cram\'er-Rao lower bound, 
for an arbitrary interconnection topology between the AP antennas, and show that the star topology is optimal. To reduce the overhead and complexity,~\cite{Chen23} decouples the calibration problem of a hybrid beamforming system into digital-chain
calibration and analog-chain calibration; the former has a closed-form solution, while the latter is solved iteratively using alternating
optimization. Argos, presented in~\cite{Shepard12Argos}, is a prototype where calibration pilot signals are exchanged with a (possibly additional)
reference antenna. To reduce the overhead of calibrating a massive array, \cite{Papadopoulos14} introduces Avalanche, a multi-stage sequential calibration method whereby,
at each stage, the already calibrated part of the array is exploited to jointly calibrate multiple antennas with a single common pilot.}

\emph{Inter}-AP calibration 
can also be performed using over-the-air signaling
between the access points. In fact, from a reciprocity calibration perspective, there is no difference in
principle between
calibrating antennas within an AP relative to one another, and calibrating different APs relative to one another
\cite{Larsson23_calibration}. However, some aspects are unique, for example, the different AP antennas
may be assigned specific roles or one can beamform the
calibration signals. Consequently, several  specific techniques for  inter-AP calibration have been developed. For example, 
\cite{Balan13} proposed AirSync, where a master AP broadcasts an out-of-band pilot and the other (slave) APs receive this pilot using a dedicated antenna. To reduce the sensitivity to antenna placement inherent in AirSync,~\cite{Rogalin14} simultaneously employs multiple anchor APs that transmit calibration signals in dedicated slots of the TDD flow. AirShare, introduced in~\cite{Abari15AirShare}, lets an emitter transmit two single frequency tones from which a
shared reference clock can be calculated. Fully distributed (no central server) calibration based on average consensus 
has been studied in~\cite{Rashid23,Rashid23b,Rashid2024}.  Considering beamformed pilots, \cite{Vieira21} sweeps the beam between all pairs of APs, while~\cite{Ganesan24_beamsync} aligns the beam with the dominant direction of the channel between APs. Reference \cite{Chen17} employs combining methods to improve the received power of calibration signals between AP clusters. Reference \cite{Kim22} proposed a gradual method consisting in calibration important APs first to reduce overhead. A cell-free massive \gls{MIMO} testbed with over-the-air calibration is presented in~\cite{Cao23}.
Recently,\cite{Larsson24_synchrony} put forth an analytical framework to analyze the error variance of estimating $\vecphi_i$ from the measurements $\vecalpha_i$ under various who-measures-on-whom topologies between the \glspl{AP}.



\subsection{\revise{Contributions}}
In this paper, we propose a framework for breaking the conventional \gls{TDD} flow to enable phase reciprocity calibration between \glspl{AP}. \revise{We assume that the transmit and receive phase noise are common to all antennas of each \gls{AP}, but do not assume any correlation between transmit and phase noise processes, and thus also address intra-AP calibration of the transmit/receive mismatches.} The contributions are summarized as follows.

\begin{itemize} 
    \item \textit{A ``broken''  \gls{TDD} flow for phase measurements:} We design a mechanism in which each \gls{AP} periodically shifts its uplink and downlink periods in selected \gls{TDD} slots, such that its downlink period partially overlaps with the uplink period of  other \glspl{AP}, and vice versa
    (see Fig.~\ref{fig:TDD_operation} for the two-AP case, to be described in detail in Section~\ref{sec:two_AP}). During these overlaps, the \gls{AP} in downlink mode transmits a phase calibration signal to the \glspl{AP} in uplink mode. Over time, this results in bidirectional measurements~$\vecalpha_i$, enabling reciprocity calibration for the subsequent downlink periods. 
    To reduce the time needed to complete a measurement of~$\vecalpha_i$, we propose an \gls{AP} scheduling scheme based on graph coloring, allowing multiple \glspl{AP} to transmit calibration signals simultaneously. 

\newlength{\UL}
\setlength{\UL}{1.5cm}

\newlength{\DL}
\setlength{\DL}{2cm}

\newlength{\guard}
\setlength{\guard}{.3cm}

\newlength{\sync}
\setlength{\sync}{.2cm}

\newlength{\pilot}
\setlength{\pilot}{.25cm}

\tikzset{uplink/.style={
draw,rectangle,
minimum height=1cm,
minimum width=\UL
}}
\tikzset{downlink/.style={
draw,rectangle,
minimum height=1cm,
minimum width=\DL
}}
    
\begin{figure}[t!]
    \centering

    \tikzset{rectangle open right/.style={
            draw=none, minimum height=1cm,
            append after command={
                [shorten <= -0.5\pgflinewidth]
                ([shift={(-1.5\pgflinewidth,-0.5\pgflinewidth)}]\tikzlastnode.north east)
            edge([shift={( 0.5\pgflinewidth,-0.5\pgflinewidth)}]\tikzlastnode.north west) 
                ([shift={( 0.5\pgflinewidth,-0.5\pgflinewidth)}]\tikzlastnode.north west)
            edge([shift={( 0.5\pgflinewidth,+0.5\pgflinewidth)}]\tikzlastnode.south west)            
                ([shift={( 0.5\pgflinewidth,+0.5\pgflinewidth)}]\tikzlastnode.south west)
            edge([shift={(-1.0\pgflinewidth,+0.5\pgflinewidth)}]\tikzlastnode.south east)
            }
        }
    }
    
    \tikzset{rectangle open left/.style={
            draw=none, minimum height=1cm,
            append after command={
                (\tikzlastnode.north west)
            edge(\tikzlastnode.north east) 
                (\tikzlastnode.north east)
            edge(\tikzlastnode.south east)            
                (\tikzlastnode.south east)
            edge(\tikzlastnode.south west)
            }
        }
    }

    \captionsetup[subfigure]{oneside,margin={0cm,0.25cm}}
    \subcaptionbox{Conventional TDD flow. Blue boxes represent uplink pilot transmissions.}{\scalebox{.83}{\begin{tikzpicture}
    \node[uplink] (ul11) at (0,0) {uplink};
    \node[rectangle open left, left=\guard of ul11] (l1) {$\!\ldots $};
    \node[left=0cm of l1] (AP1) {$\!$AP $1$};
    \node[downlink,right=\guard of ul11] (dl11) {downlink};
    \node[uplink,right=\guard of dl11] (ul12) {uplink};
    \node[downlink,right=\guard of ul12] (dl12) {downlink};
    \node[rectangle open right, right=\guard of dl12] (r1) {$ \ldots\!$};

    \node[uplink] (ul21) at (0,-1.5cm) {uplink};
    \node[rectangle open left, left=\guard of ul21] (l2) {$\!\ldots $};
    \node[left=0cm of l2] (AP2) {$\!$AP~$2$};
    \node[downlink,right=\guard of ul21] (dl21) {downlink};
    \node[uplink,right=\guard of dl21] (ul22) {uplink};
    \node[downlink,right=\guard of ul22] (dl22) {downlink};
    \node[rectangle open right, right=\guard of dl22] () {$ \ldots\!$};

    \draw[gray!50] ([yshift=.5cm]ul11.north west) -- ([yshift=-1.5cm]ul11.south west);
    \draw[gray!50] ([yshift=.5cm]ul12.north west) -- ([yshift=-1.5cm]ul12.south west);
    \draw[gray!50] ([yshift=.5cm]r1.north west) -- ([yshift=-1.5cm]r1.south west);

    \fill[blue!70,opacity=.5] (ul11.north west) rectangle ([xshift=\pilot]ul11.south west);
    \fill[blue!70,opacity=.5] (ul21.north west) rectangle ([xshift=\pilot]ul21.south west);
    \fill[blue!70,opacity=.5] (ul12.north west) rectangle ([xshift=\pilot]ul12.south west);
    \fill[blue!70,opacity=.5] (ul22.north west) rectangle ([xshift=\pilot]ul22.south west);
    
    \draw[<->] ([yshift=.3cm]ul11.north west) -- node[midway,above] () {$\tc$} ([yshift=.3cm]ul12.north west);
\end{tikzpicture}
} \label{fig:TDD}} 

    \smallskip
    \captionsetup[subfigure]{oneside,margin={0cm,0.25cm}}
    \subcaptionbox{\revise{A calibration scheme in which, once every $\nsync$ slots, a dedicated section of the downlink is reserved for phase measurements. This is done in AirSync~\cite{Balan13} and~\cite{Rogalin14}. Orange boxes represent the instants of phase measurements, with the arrows showing the direction of calibration signals.}
    }{\scalebox{.83}{\begin{tikzpicture}
    \node[uplink] (ul11) at (0,0) {uplink};
    \node[rectangle open left, left=\guard of ul11] (l1) {$\!\ldots$};
    \node[left=0cm of l1] (AP1) {$\!$AP~$1$};
    \node[draw=black,minimum height=1cm,
minimum width=\DL-2*\guard - 2*\sync,rectangle,right=3*\guard + 2*\sync of ul11,align=center] (dl11) {$\!$down-$\!$\\link};
    \node[uplink,right=\guard of dl11] (ul12) {uplink};
    \node[downlink,right=\guard of ul12] (dl12) {downlink};
    \node[rectangle open right, right=\guard of dl12] (r1) {$\ldots\!$};

    \node[uplink] (ul21) at (0,-1.5cm) {$\!$uplink$\!$};
    \node[rectangle open left, left=\guard of ul21] (l2) {$\!\ldots$};
    \node[left=0cm of l2] (AP2) {$\!$AP~$2$};
    \node[draw=black,minimum height=1cm,
minimum width=\DL-2*\guard - 2*\sync,rectangle,right=3*\guard + 2*\sync of ul21,align=center] (dl21) {$\!$down-$\!$\\link};
    \node[uplink,right=\guard of dl21] (ul22) {uplink};
    \node[downlink,right=\guard of ul22] (dl22) {downlink};
    \node[rectangle open right, right=\guard of dl22] () {$\ldots\!$};

    \draw[gray!70] ([yshift=.5cm]ul11.north west) -- ([yshift=-1.5cm]ul11.south west);
    \draw[gray!70] ([yshift=.5cm]ul12.north west) -- ([yshift=-1.5cm]ul12.south west);
    \draw[gray!70] ([yshift=.5cm]r1.north west) -- ([yshift=-1.5cm]r1.south west);

    \fill[blue!80,opacity=.5] (ul11.north west) rectangle ([xshift=\pilot]ul11.south west);
    \fill[blue!80,opacity=.5] (ul21.north west) rectangle ([xshift=\pilot]ul21.south west);
    \fill[blue!80,opacity=.5] (ul12.north west) rectangle ([xshift=\pilot]ul12.south west);
    \fill[blue!80,opacity=.5] (ul22.north west) rectangle ([xshift=\pilot]ul22.south west);
    

    \fill[orange, draw=black] ([xshift=\guard]ul11.south east) rectangle ([xshift=\sync+\guard]ul11.north east);
    \fill[orange, draw=black] ([xshift=\sync+2*\guard]ul11.south east) rectangle ([xshift=2*\sync+2*\guard]ul11.north east);

    \fill[orange, draw=black] ([xshift=\guard]ul21.south east) rectangle ([xshift=\sync+\guard]ul21.north east);
    \fill[orange, draw=black] ([xshift=\sync+2*\guard]ul21.south east) rectangle ([xshift=2*\sync+2*\guard]ul21.north east);

    \draw[-latex] ([xshift=\guard+0.5*\sync,yshift=.5cm]ul21.south east) -- ([xshift=\guard+0.5*\sync,yshift=.5cm]ul11.south east);

    \draw[-latex] ([xshift=2*\guard+1.5*\sync,yshift=.5cm]ul11.south east) -- ([xshift=2*\guard+1.5*\sync,yshift=.5cm]ul21.south east);

    \draw[->] ([xshift=-2*\guard,yshift=-.35cm]ul21.south east) -- node[below=.1cm,pos=0,align = center] () {measure \\ $\nu_{1,i_1}- \nu_{2,i_1}$} ([xshift=\guard+0.5*\sync,yshift=-.05cm]ul21.south east);

    \draw[->]([xshift=12*\guard,yshift=-.35cm]ul21.south west) -- node[below=.1cm,pos=0,align = center] () {measure \\ $\nu_{2,i_2} - \nu_{1,i_2}$} ([xshift=2*\guard+1.5*\sync,yshift=-.05cm]ul21.south east);

    \draw[<->] ([yshift=.3cm]ul11.north west) -- node[midway,above] () {$\tc$} ([yshift=.3cm]ul12.north west);
\end{tikzpicture}
}\label{fig:TDD_dedicated_sync_slot}}

    \smallskip
    \captionsetup[subfigure]{oneside,margin={0cm,0.25cm}}
    \subcaptionbox{Proposed ``broken'' TDD flow: once every $\nsync$ slots, \gls{AP}~2 moves the last $1 + \tg$ samples of the uplink to the end of the slot, and shifts the downlink earlier accordingly. 
    }{\scalebox{.83}{\begin{tikzpicture}
    \node[uplink] (ul11) at (0,0) {uplink};
    \node[rectangle open left, left=\guard of ul11] (l1) {$\!\ldots$};
    \node[left=0cm of l1] (AP1) {$\!$AP~$1$};
    \node[downlink,right=\guard of ul11] (dl11) {downlink};
    \node[uplink,right=\guard of dl11] (ul12) {uplink};
    \node[downlink,right=\guard of ul12] (dl12) {downlink};
    \node[rectangle open right, right=\guard of dl12] (r1) {$\ldots\!$};

    \node[uplink,minimum width=\UL-\guard-\sync] (ul21) at (-0.5*\guard-0.5*\sync,-1.5cm) {$\!$uplink$\!$};
    \node[rectangle open left, left=\guard of ul21] (l2) {$\!\ldots$};
    \node[left=0cm of l2] (AP2) {$\!$AP~$2$};
    \node[downlink,right=\guard of ul21] (dl21) {downlink};
    \node[uplink,minimum width=\UL+\guard+\sync,right=\guard of dl21] (ul22) {uplink};
    \node[downlink,right=\guard of ul22] (dl22) {downlink};
    \node[rectangle open right, right=\guard of dl22] () {$\ldots\!$};

    \draw[gray!70] ([yshift=.5cm]ul11.north west) -- ([yshift=-1.5cm]ul11.south west);
    \draw[gray!70] ([yshift=.5cm]ul12.north west) -- ([yshift=-1.5cm]ul12.south west);
    \draw[gray!70] ([yshift=.5cm]r1.north west) -- ([yshift=-1.5cm]r1.south west);

    \fill[blue!80,opacity=.5] (ul11.north west) rectangle ([xshift=\pilot]ul11.south west);
    \fill[blue!80,opacity=.5] (ul21.north west) rectangle ([xshift=\pilot]ul21.south west);
    \fill[blue!80,opacity=.5] (ul12.north west) rectangle ([xshift=\pilot]ul12.south west);
    \fill[blue!80,opacity=.5] ([xshift=\guard+\sync]ul22.north west) rectangle ([xshift=\pilot+\guard+\sync]ul22.south west);
    
    \fill[orange] (dl11.south east) rectangle ([xshift=-\sync]dl11.north east);

    \fill[orange] (ul11.south east) rectangle ([xshift=-\sync]ul11.north east);

    \fill[orange] (dl21.south west) rectangle ([xshift=\sync]dl21.north west);
    \fill[orange] (ul22.south west) rectangle ([xshift=\sync]ul22.north west);

    \draw[-latex] ([xshift=-0.5*\sync,yshift=.5cm]dl11.south east) -- ([xshift=0.5*\sync,yshift=.5cm]ul22.south west);

    \draw[-latex] ([xshift=0.5*\sync,yshift=.5cm]dl21.south west) -- ([xshift=-0.5*\sync,yshift=.5cm]ul11.south east);

    \draw[->] ([xshift=0.5*\sync,yshift=-.35cm]dl21.south west) -- node[below=.1cm,near start,align = center] () {time $i_1$, \\ measure \\ $\nu_{1,i_1}- \nu_{2,i_1}$} ([xshift=0.5*\sync,yshift=-.05cm]dl21.south west);

    \draw[->]([xshift=0.5*\sync,yshift=-.35cm]ul22.south west) -- node[below=.1cm,near start,align = center] () {time $i_2$, \\  measure \\ $~~~\nu_{2,i_2} - \nu_{1,i_2}$} ([xshift=0.5*\sync,yshift=-.05cm]ul22.south west);

    \draw[<->] ([yshift=.3cm]ul11.north west) -- node[midway,above] () {$\tc$} ([yshift=.3cm]ul12.north west);
\end{tikzpicture}
}\label{fig:shifted_TDD}}
    \caption{Illustration of two consecutive slots following the conventional or proposed ``broken'' \gls{TDD} operation. 
    } 
    
    \label{fig:TDD_operation}
\end{figure}

    \item \textit{A Kalman filter for phase tracking:} To enhance the phase estimation accuracy, we develop a Kalman filter for tracking the evolution of $\vecalpha_i$ over time. The filter accounts for both measurement noise and phase drift. From the filter outputs, we estimate $\vecphi_i$ using a least-squares method. 
    
    \item \textit{\Gls{SE} analysis:} We derive a downlink \gls{SE} achievable with coherent conjugate beamforming and \gls{ZF} beamforming after the estimate of $\vecphi_i$ is used for phase calibration. The \gls{SE} formulas allow for pinpointing the impact of residual phase noise.

    \item \textit{Insights from numerical experiments:} 
    We numerically evaluate the achievable \revise{per-\gls{UE} downlink} \gls{SE} for typical network configurations, channel conditions, and LOs' qualities. \revise{The \gls{SE} achieved with our proposed scheme outperforms schemes inspired by AirSync~\cite{Balan13} and \cite{Rogalin14} where calibration signals are transmitted in dedicated slots.} The results \revise{also} reveal several important insights. First, frequent phase reestimation is critical for coherent beamforming, particularly when the \glspl{LO} exhibit rapid phase drift. 
    Second, using more APs initially enhances SE due to spatial diversity, but also increases calibration overhead and residual phase noise. 
    Finally, we show that denser who-measures-on-whom topologies between the APs 
    can degrade the SE when the phase drifts during the measurement duration and power sharing between calibration signals are accounted for. These results highlight the importance of balancing measurement frequency, AP interconnection density, and downlink service continuity.
\end{itemize}

Among the related works mentioned in Section~\ref{sec:relatedwork}, only~\cite{Rogalin14} addresses the placement of phase measurements within a TDD flow. However, pilot signals are transmitted in dedicated slots, rather than interleaved into data-carrying slots as in our paper. \revise{ Note that our scheme
shifts the uplink–downlink switching instants in the TDD frame but does
not increase the number of such instants. In contrast, schemes that introduce dedicated calibration slots add extra switching instants within those
slots, which incurs significant overhead after the required guard intervals
are accounted for. } Furthermore, the LO drift is not considered in~\cite{Rogalin14}. The impact of phase noise on downlink beamforming was investigated for co-located massive MIMO in~\cite{Krishnan2016}. However, this analysis assumes that the transmit and receive phase noise are identical (contrary to our model where they have opposite signs, consistent with~\cite{Nissel22}), and thus neglects the need for phase measurements between the base station antennas.


\subsection{\revise{Paper Organization and} Notation}
\revise{The remainder of the paper is organized as follows. In Section~\ref{sec:model}, we present the system model, including the reciprocity calibration scheme assuming known $\vecphi_i$. Next, we propose the broken \gls{TDD} flow to obtain $\vecphi_i$, presenting the two-\gls{AP} case in Section~\ref{sec:two_AP} and the general case in Section~\ref{sec:general_case}. In Section~\ref{sec:rate}, we derive the achievable \gls{SE} of the proposed scheme. Numerical results and discussions are given in Section~\ref{sec:result}. Finally, we conclude the paper in Section~\ref{sec:conclusion}.  Mathematical preliminaries, proofs, and detailed calculations are deferred to the appendices.}

	We denote scalars with plain italic letters, e.g., $x$ and $X$, 
    column vectors with lowercase boldface letters, e.g., $\vecx$, 
    and matrices with uppercase boldface letters, e.g., $\matX$. 
    The superscripts~$^*$, $\tp{}$, and~$\herm{}$ denote the conjugate, transpose, and conjugate transpose, respectively.
	By $\matidentity_m$, $\vecone_m$, and $\veczero_m$, we denote the $m\times m$ 
    identity matrix, $m\times 1$ all-one vector, and $m\times 1$ all-zero vector, respectively. 
    We denote the set of integers from $m$ to $n$ by $[m:n]$; $[n] = [1:n]$. 
    For $\vecx = \tp{[x_1 \dots x_n]}$, we use $e^\vecx$ to denote $\tp{[e^{x_1} \dots e^{x_n}]}$, and $\matD_{\vecx}$ the diagonal matrix with the main diagonal given by~$\vecx$. We denote the imaginary unit by $\jmath$ and the argument of the complex number~$x$ by $\angle x$. All phase values are defined mod $2\pi$.
    
\section{System Model} \label{sec:model}

We consider a system with $L$ \glspl{AP}, each equipped with $N$
antennas, serving $K$ single-antenna \glspl{UE}. The \glspl{AP} are
connected to a \gls{CPU} via fronthaul links. Each \gls{AP} is driven
by a single, independent~\gls{LO}.

\subsection{Channel and Phase Noise Model}
We consider a narrowband channel where samples are taken sequentially in time without intersymbol interference. We also use ``sample'' to refer to a discretized time unit. Let $\vech_{k,\ell} \in \complexset^{N}$ denote the channel vector between \gls{UE}~$k\in [K]$ and \gls{AP}~$\ell\in [L]$. 
We assume a block fading model where $\vech_{k,\ell}$ remains constant for each {time} block of~$\tc$ samples 
and varies independently between blocks.\footnote{The block fading assumption is used to simplify the signal model, but is not conceptually necessary for the proposed phase calibration method.}
Furthermore, we consider \gls{iid} Rayleigh fading with $\vech_{k,\ell} \sim \jpg(\veczero, \beta_{k,\ell}\matidentity_N)$, where $\beta_{k,\ell}$ is the large-scale fading coefficient. It will make no material difference if another channel model, e.g., correlated Rayleigh fading, is considered.
For $i,j \in [L]$, let $\matG_{i,j} \in \complexset^{N \times N}$ be the propagation channel from \gls{AP}~$i$ to \gls{AP}~$j$. This channel is reciprocal, i.e., $\matG_{j, i} = \tp{\matG}_{i, j}$. We assume that each AP~$i$ knows $\matG_{j,i}$, for all $j \ne {i}$.\footnote{We shall see that it is enough for AP~$i$ and AP~$j$ to know $\matG_{i,j}$ up to a common phase shift.} 
This assumption is reasonable when the \glspl{AP} are static and thus their slowly varying channels can be tracked accurately. 



We assume that phase noise at the \glspl{UE} is perfectly tracked with
sufficiently frequent demodulation pilots, which we do not model.\footnote{\revise{Once the APs are calibrated, demodulation pilots are incorporated in the downlink transmission in the conventional manner and do not affect the proposed AP-side calibration procedure.}} 
We focus on phase noise at the \glspl{AP}. The
free-running \gls{LO} at \gls{AP}~$\ell$ induces \revise{a transmit phase noise
$\nut_{\ell,i}$ and a receive phase noise $\nur_{\ell,i}$, both} common to all antennas,\footnote{Calibrating the
antennas \emph{within an AP} can be done much more   infrequently using bidirectional intra-AP measurements, see Section~\ref{sec:relatedwork}.}
at sample time~$i$.  The resulting
multiplicative transmit and receive phase noise are
\revise{$\exp(-\jmath \nut_{\ell,i})$ and $\exp(\jmath \nur_{\ell,i})$},
respectively. Here, following the convention
in~\cite{Larsson23_calibration,Larsson24_synchrony}, and consistent with~\cite{Nissel22}, we use opposite
signs for the transmit and receive phase noise, as transmitted and
received signals travel in opposite directions. Furthermore, we
assume that 
\revise{the random processes~$\nut_{\ell,i}$and~$\nur_{\ell,i}$ evolve over time.} 

\subsection{Signal Model}
We consider \gls{TDD} operation. Each \gls{TDD} slot corresponds to a length-$\tc$ coherence block, and is divided into different periods: 1) $\tpl$ samples for uplink pilot transmission, 2) $\tu$ samples for uplink data transmission, 3) $\td$ samples for downlink data transmission, and 4) two guard periods of $\tg$ samples each that separate the uplink and downlink periods. During both uplink pilot transmission and uplink data transmission, we say that the \gls{AP} is in uplink mode. In this paper, we focus on the analysis of the downlink data transmission. Note that LO phase noise is not a problem for the uplink, since on uplink pilots and data see the same channel.
We next describe the uplink pilot transmission and downlink data transmission periods without specifying their positions within a slot. 

\subsubsection{Uplink Pilot Transmission}

We assume that \revise{$\tpl \ge K$} and let 
user~$k$ transmit {the known pilot sequence $\sqrt{\Ppl \revise{\tpl}} \vece_{\tpl,k}$, where $\vece_{\tpl,k}$ is the length-$\revise{\tpl}$ canonical basis vector with a one at position~$k$ and zeros elsewhere.}\footnote{Note that pilots can be spread out during the uplink period. However, for the convenience of tracking the time indices (which is important for tracking the phase noise evolution), we assume that uplink pilots are transmitted at the beginning of the slot. Furthermore, we assume no pilot contamination (by considering \revise{$\tpl \ge K$}) to focus on the impact of phase noise.}
The received signal over the pilot period correlated with $\vece_k$ equals the received signal at {sample} time $k$ of the current slot, and is given by
\begin{equation}
    \vecy_{k,\ell}\supp{pilot} = \sqrt{\Ppl \revise{\tpl}} \exp(\jmath \revise{\nur_{\ell,k}}) \vech_{k,\ell} + \vecz\supp{pilot}_{k,\ell}
\end{equation}
where $\vecz\supp{pilot}_{k,\ell} \sim \jpg(\veczero_N, \matidentity_N)$ is the \gls{AWGN}. Throughout the paper, we normalize so that the noise components have unit variance. 
Note that the subscript~$k$ in \revise{$\nur_{\ell,k}$} refers to sample time $k$
of the current slot. As \revise{$\nur_{\ell,k}$} is uniform over $[-\pi,\pi]$, only the effective channel vector $\vecq_{k,\ell} = \exp(\jmath \revise{\nur_{\ell,k}}) \vech_{k,\ell}$ can be estimated. Its \gls{LMMSE} estimate is given by
\begin{equation}
    \hat \vecq_{k,\ell} = c_{k,\ell} \vecy_{k,\ell}\supp{pilot} 
\end{equation}
with $c_{k,\ell} = \dfrac{\sqrt{\Ppl \revise{\tpl}} \beta_{k,\ell}}{\!\Ppl \revise{\tpl} \beta_{k,\ell} +\! 1}$. It holds that $\hat \vecq_{k,\ell} \sim \jpg\left(\veczero, \gamma_{k,\ell}\matidentity_N\right)$  with $\gamma_{k,\ell} = \sqrt{\Ppl \revise{\tpl}} \beta_{k,\ell} c_{k,\ell}$. We denote $\vecgamma_\ell = \tp{[\gamma_{1,\ell}\ \dots \ \gamma_{K,\ell}]}$.



\subsubsection{Downlink Data Transmission}
\label{sec:downlink}
Consider downlink transmission at sample~$i$. Here, we let the index $i$ be incremented across slots, i.e., $i$ can exceed $\tc$. We allow for the possibility that only one \gls{AP} is in downlink mode at a given time, and let $a_{\ell,i} \in \{0,1\}$ indicate if AP~$\ell$ is in downlink period ($a_{\ell,i} = 1$) or not ($a_{\ell,i} = 0$) at time~$i$. 
Specifically, \gls{AP}~$\ell$ transmits 
    \begin{align}
    \vecx_{\ell,i} &= a_{\ell,i} 
    \sqrt{\Pdl} \sum_{k=1}^K \sqrt{\eta_{k,\ell}} \vecw_{k,\ell,i} s_{k,i} \\
    &= a_{\ell,i} 
    \sqrt{\Pdl} \matW_{\ell,i} \matD_{\veceta_\ell}^{1/2} \vecs_{i}, \label{eq:tmp330}
\end{align}
where $\vecw_{k,\ell,i}$ is the beamforming vector of AP~$\ell$ for \gls{UE}~$k$, with unit norm on average; $s_{k,i}$, with $\Exop[|s_{k,i}|^2] = 1$, is the data signal intended for \gls{UE}~$k$; 
$\eta_{k,\ell}$, $\ell \in \{1,2\}$, $k \in [K]$, are the power control coefficients chosen to satisfy the power constraint $\Exop[\|\vecx_{\ell,i}\|^2] \le \Pdl$, i.e., 
$\sum_{k=1}^K \eta_{k,\ell} \le 1$, $\ell \in [L]$. Furthermore, in~\eqref{eq:tmp330}, $\matW_{\ell,i}= [\vecw_{1,\ell,i} \dots \vecw_{K,\ell,i}]$ 
and $\veceta_\ell = \tp{[\eta_{1,\ell}\ \dots \ \eta_{K,\ell}]}$. The beamforming vectors are designed based on $\widehat \matQ_{\ell,i} = [\hat \vecq_{1,\ell,i} \dots \hat \vecq_{K,\ell,i}] \in \complexset^{N \times K}$, where $\hat \vecq_{k,\ell,i}$ is the latest estimate of $\vecq_{k,\ell}$ up to time $i$. Specifically, the effective channel estimated by $\hat \vecq_{k,\ell,i}$ is $\vecq_{k,\ell,i} = \exp(\jmath \revise{\nur_{\ell, [i]_k}}) \vech_{k,\ell}$ where
\begin{equation}
    [i]_k = i - 1 - [(i - 1 - k) \mod \tc]    
\end{equation}
is the index of the sample where \gls{UE}~$k$ transmitted its latest pilot. The index $[i]_k$ simply refers to the $k$th sample of the slot that contains $i$.
We consider two types of beamforming: 

\begin{itemize}[leftmargin = *]
    \item \textit{Conjugate beamforming} with 
    $\vecw_{k,\ell,i} = (N \gamma_{k,\ell})^{-1/2}\hat \vecq_{k,\ell,i}^*$, 
    i.e., $\matW_{\ell,i} = \sqrt{N}\widehat \matQ_{\ell,i}^* \matD_{\vecgamma_\ell}^{-1/2}$. 
    \gls{UE}~$k$ receives
    \begin{align}
        y_{k,i} &= \sum_{\ell=1}^L \exp({-\jmath \revise{\nut_{\ell,i}}}) \tp{\vech}_{k,\ell} \vecx_{\ell,i} + z_{k,i} \\
        &= \sqrt{\Pdl}\sum_{\ell = 1}^L a_{\ell,i} \exp(-\jmath\revise{\nut_{\ell,i}}) 
        \tp{\vech}_{k,\ell} \notag \\
        &\quad \cdot \sum_{k'=1}^K \sqrt{\frac{\eta_{k',\ell}}{N\gamma_{k',\ell}}}\hat\vecq_{k',\ell,i}^* s_{k',i} + z_{k,i} \\
        &= \sqrt{\Pdl}\sum_{\ell = 1}^La_{\ell,i} \sum_{k'=1}^K\sqrt{\frac{\eta_{k',\ell}}{N\gamma_{k',\ell}}} \exp[-\jmath (\revise{\nut_{\ell,i} +\nur_{\ell,[i]_k}}) 
        ] \notag \\
        &\quad \cdot \tp{\vecq}_{k,\ell,i} \hat\vecq_{k',\ell,i}^* s_{k',i}  + z_{k,i} \label{eq:signal_dl_conj} 
    \end{align}
    where $z_{k,i} \sim \jpg(0,1)$ is the \gls{AWGN}. Notice that, because of the phase drift between the time when the channel is estimated and when it is used for beamforming, \revise{the two phase terms $\nut_{\ell,i}$ and $\nur_{\ell,[i]_k}$ have different time indices. 
    In the absence of phase drift, they appear as $\nut_{\ell}$ and $\nur_{\ell}$, as in~\cite{Larsson23_calibration,Larsson24_synchrony}.}
    
    \item \textit{\Gls{ZF} beamforming} with 
    \begin{equation}
        \matW_{\ell,i} = \sqrt{N-K} \widehat \matQ_{\ell,i}^*\big(\tp{\widehat{\matQ}}_{\ell,i} \widehat \matQ_{\ell,i}^*\big)^{-1} \matD_{\vecgamma_\ell}^{1/2}.    
    \end{equation}
    Whenever ZF beamforming is used, we assume that $N \ge K$. 
    The collective received signal across \glspl{UE} is given by
    \begin{align}
        \vecy_{i} = \tp{[y_{1,i} \dots y_{K,i}]} = \sum_{\ell = 1}^L \exp({-\jmath \revise{\nut_{\ell,i}}}) \tp{\matH}_{\ell} \vecx_{\ell,i} + \vecz_{i} 
    \end{align}
    where $\matH_\ell = [\vech_{1,\ell} \dots \vech_{K,\ell}]$. This matrix can be expressed as $\matH_\ell = \matQ_{\ell,i} \matV_{\ell,i}$ with $\matQ_{\ell,i} = [\vecq_{1,\ell,i} \dots \vecq_{K,\ell,i}]$ and $\matV_{\ell,i}$ being the diagonal matrix with the diagonal $(\exp(-\jmath \revise{\nur_{\ell,[i]_1}}) \dots \exp(-\jmath \revise{\nur_{\ell,[i]_K}}))$. We denote the channel estimation error as $\tilde{\vecq}_{k,\ell,i} = \vecq_{k,\ell,i} - \hat{\vecq}_{k,\ell,i}$ and  $\widetilde \matQ_{\ell,i} = [\tilde \vecq_{1,\ell,i} \dots \tilde \vecq_{K,\ell,i}]$. By writing $\matH_\ell = (\widehat \matQ_{\ell,i} + \widetilde \matQ_{\ell,i}) \matV_{\ell,i}$ and using $\tp{\widehat \matQ}_{\ell,i}  \matW_{\ell,i} = \sqrt{N-K}\matD_{\vecgamma_\ell}^{1/2}$, we further expand $\vecy_{i}$ as
    \begin{align}
        \vecy_{i} = &\sqrt{(N-K)\Pdl} \sum_{\ell = 1}^L a_{\ell,i} \exp(-\jmath \revise{\nut_{\ell,i}}) 
        \matV_{\ell,i} \matD_{\vecgamma_\ell}^{1/2} \matD_{\veceta_\ell}^{1/2} \vecs_{i} \notag \\
         &+ \sum_{\ell = 1}^L \! \exp(-\jmath \revise{\nut_{\ell,i}}) \matV_{\ell,i} \tp{\widetilde \matQ}_{\ell,i}\vecx_{\ell,i} 
        + \vecz_{i}.\!
    \end{align}
    It follows that the signal received by \gls{UE}~$k$ is given by
    \begin{align}
        y_{k,i} &= \sqrt{(N-K)\Pdl} \sum_{\ell = 1}^L a_{\ell,i} \sqrt{\eta_{k,\ell} \gamma_{k,\ell}} \notag \\
        &\qquad \cdot \exp[-\jmath (\revise{\nut_{\ell,i} + \nur_{\ell,[i]_k}}) 
        ] s_{k,i} \notag \\
        &\quad + \sum_{\ell = 1}^L \exp[-\jmath (\revise{\nut_{\ell,i} + \nur_{\ell,[i]_k}})] \tp{\tilde \vecq}_{k,\ell,i}\vecx_{\ell,i} 
        + z_{k,i}. \label{eq:signal_dl_ZF} 
    \end{align}
\end{itemize}


\subsection{Phase Noise Compensation} \label{sec:PN_compensation}

For both conjugate beamforming and \gls{ZF} beamforming, in the received signal $ y_{k,i}$, 
the desired signal $s_{k,i}$ is corrupted by the phase noise $\revise{-\nut_{\ell,i} - \nur_{\ell,[i]_k}}$. To reduce complexity, we ignore the difference between $\{[i]_k\}_{k\in [K]}$ and represent all $\{\revise{\nut_{\ell,i} + \nur_{\ell,[i]_k}}\}_{k \in [K]}$ by\footnote{\revise{This allows us to track a common phase-compensation term rather than  \gls{UE}-specific ones. It induces a mismatch due to the phase drift between $[i]_k$ and $[i]_{\floor{K/2}}$, which we shall account for in the Kalman filter design.}} 
\begin{equation}
    \phi\of{\ell}_i = 
    \revise{\nut_{\ell,i} +\nur_{\ell,[i]_{\floor{K/2}}}}. \label{eq:omega_ell}
\end{equation} 
To compensate for this phase noise, we let \gls{AP}~$\ell$ multiply its downlink signal $\vecx_{\ell,i}$ by $\exp(\jmath \theta_{\ell,i})$ 
with
\begin{equation}
     \theta_{\ell,i} =  
     \phi\of{\ell}_i  + c 
     \label{eq:manyAP_theta}
\end{equation}
where $c$ is an arbitrary constant common to all \glspl{AP}. Note that bidirectional measurements between the \glspl{AP} only allow to estimate pairwise differences between $\phi_i\of{\ell}$ across $\ell$. Therefore, $\phi_i\of{\ell}$ can only be estimated up to a common additive constant, which suffices~\eqref{eq:manyAP_theta}.
\gls{UE}~$k$ sees the phase shift 
\begin{equation}
    \revise{-\nut_{\ell,i} - \nur_{\ell,[i]_k}} + \theta_{\ell,i} = \revise{\nur_{\ell,[i]_{\floor{K/2}}} - \nur_{\ell,[i]_k}} + c \approx c,
\end{equation} 
and compensate for it by estimating $c$ via a downlink demodulation pilot. As phase estimation via demodulation pilot is well understood, we omit its details. 
%
%
Next, we propose a scheme for the \glspl{AP} to estimate $\{\phi\of{\ell}_i\}_{\ell \in [L]}$, up to a common additive constant, via bidirectional measurements. 

    

\section{Breaking TDD for Phase Calibration: \\The Two-AP Case}\label{sec:two_AP}

We first focus on the two-\gls{AP} case. 
We let $c = - \phi\of{1}_i$. Our strategy is to estimate $\phi\of{2}_i - \phi\of{1}_i$ 
at periodic times $i$, and reset the phase compensation term $\theta_{\ell,i'}$, 
$\forall i' > i$, 
as in~\eqref{eq:manyAP_theta} 
whenever a new estimate is obtained.\footnote{In phase noise processes characterized by independent increments, such as the Wiener process, the \gls{MMSE} predictor of future values is given by the current estimate, as the future is statistically independent of the past.}
To estimate $\phi\of{2}_i - \phi\of{1}_i$, 
we modify the TDD flow to accommodate the transmission of phase calibration signals as follows. Starting with the conventional \gls{TDD} flow shown in Fig.~\ref{fig:TDD_operation}(a), we group $F$ slots into a frame and let \gls{AP}~2 shift the periods within the first slot of each frame 
as illustrated in Fig.~\ref{fig:TDD_operation}(b). Specifically, AP~2 moves the last $\tg+1$ samples of the uplink period to the end of the slot, and shifts the downlink earlier accordingly. This way, the uplink of AP~1 has one overlapping sample with the downlink of AP~2, and vice versa. During these overlapping samples, the downlink AP can transmit a phase calibration signal to the uplink AP. 
This broken TDD pattern is communicated to the \glspl{UE} with negligible cost.



\subsection{Phase Measurement} \label{sec:phase_measurement}
The indices of samples where phase calibration signals are sent in the current slot are 
\begin{equation}
    i_1 = \tpl + \tu, \quad i_2 = \tpl + \tu + \tg + \td,
\end{equation}
as indicated in Fig.~\ref{fig:TDD_operation}(b).
At time $i_1$, \gls{AP}~$1$ is in uplink, and \gls{AP}~$2$ transmits the calibration signal
    $\vecx\supp{sync}_{2} = 
    \sqrt{\Pdl} \vecu_2$
where $\vecu_2$ is a unit-norm vector.
The received signal at \gls{AP}~$1$ is\footnote{\revise{For notational simplicity, we assume here that the calibration signals are transmitted with the same power as the downlink data. In numerical experiments, we shall consider also the case where the transmit power of calibration signals is reduced.}} 
\begin{align}
    \vecy_{1,i_1}\supp{sync} &= \sqrt{\Pdl}\exp\big(\jmath \alpha\of{2 \to 1}_{i_1}\big) \matG 
    \vecu_2 
    + \vecz_{1,i_1}\supp{sync} 
\end{align}
where $\alpha\of{2 \to 1}_{i_1} =  \revise{-\nut_{2,i_1} + \nur_{1,i_1}}$. \gls{AP}~$1$ estimates $\alpha\of{2 \to 1}_{i_1}$ by
$\bar \alpha\of{2 \to 1}_{i_1} = \angle \herm{\vecu}_2 \herm{\matG} \vecy_{1,i_1}\supp{sync}$. Similarly, at time $i_2$, \gls{AP}~$1$ transmits the calibration signal
    $\vecx\supp{sync}_{1} = 
    \sqrt{\Pdl} \vecu_1$ where $\vecu_1$ is a unit-norm vector.
\gls{AP}~$2$ receives
\begin{equation}
    \vecy_{2,i_2}\supp{sync} = \sqrt{\Pdl}\exp\big(\jmath \alpha\of{1 \to 2}_{i_2}\big) \tp{\matG} \vecu_1 
     + \vecz_{2,i_2}\supp{sync} 
\end{equation}
with $\alpha\of{1 \to 2}_{i_2} =  \revise{-\nut_{1,i_2} + \nur_{2,i_2}}$, and estimates $\alpha\of{1 \to 2}_{i_2}$ by
$\bar \alpha\of{1 \to 2}_{i_2} = \angle \herm{\vecu}_1 \matG^* \vecy_{2,i_2}\supp{sync}$. During times $i_1$ and $i_2$, we assume that no data is transmitted. The \glspl{AP} send $\bar \alpha\of{2\to 1}_{i_1}$ and $\bar \alpha\of{1\to 2}_{i_2}$ to the \gls{CPU}, which 
uses
\begin{equation}
    \bar \alpha_{i_2} = \bar \alpha\of{1\to 2}_{i_2} - \bar \alpha\of{2\to 1}_{i_1}  \label{eq:2AP_phi_hat} 
\end{equation}
as an estimate of 
\revise{
\begin{equation}
    \alpha\of{1\to 2}_{i_2} -  \alpha\of{2\to 1}_{i_1} = (\nut_{2,i_2} + \nur_{2,i_1}) - (\nur_{1,i_2} + \nut_{1,i_1}). \label{eq:tmp705}
\end{equation}
} 
\revise{We remark that it is enough for the \glspl{AP} to know $\matG$ up to a common phase shift. Such phase shift appears in both $\bar\alpha_{i1}\of{2\to 1}$ and $\bar\alpha_{i2}\of{1\to 2}$, and thus cancel out in $\bar{\alpha}_{i2}$. In fact, the beamforming and combining vectors can also be designed without the knowledge of $\matG$, as long as they result in a common phase term in $\bar{\alpha}_{i2}\of{1\to 2}$ and $\bar{\alpha}_{i1}\of{2\to 1}$.  Furthermore, this method also works when the calibration signals are transmitted without beamforming, i.e., for single-antenna \glspl{AP}\textemdash see~\cite[Section III-B]{Larsson23_calibration}.}
%

Following an analysis similar to~\cite[Sec.~III-B]{Ganesan24_beamsync}, we set $\vecu_1$ and~$\vecu_2$ as the leading left- and leading right-singular vectors of~$\matG$, respectively. 
The \gls{MSE} of the estimation of both $\alpha\of{2\to 1}_{i_1}$ and $\alpha\of{1\to 2}_{i_2}$ can be approximated as $\frac12 \Pdl^{-1}\opnorm{\matG}^{-2}$, where $\opnorm{\matG}$ is the operator norm 
of $\matG$. 

    We remark that the two APs only need to know $\matG$ up to a common phase shift, because 
    this phase shift is present in both $\alpha\of{2\to 1}_{i_1}$ and $\alpha\of{1\to 2}_{i_2}$, and thus canceled in their difference.  See~\cite[Appendix]{Ganesan24_beamsync} for a detailed explanation.
While the \gls{CPU} can directly use 
$\bar \alpha_{i_2}$
as an estimate of $\phi\of{2}_{i_2} - \phi\of{1}_{i_2}$ (ignoring the time offset between $i_1$ and $[i_2]_{\floor{K/2}}$), 
 the estimation of $\phi\of{2}_i - \phi\of{1}_i$ can be improved using a Kalman filter that we present next. 

\subsection{Kalman Filter} \label{sec:kalman}
To make the design of the Kalman filter explicit, we consider the discrete-time Wiener phase noise \revise{processes}\footnote{Our method is conceptually applicable to other phase noise processes where, e.g, the innovations have non-Gaussian statistics. \revise{We note that the Wiener phase noise model is reasonable if LOs are not locked to a reference, i.e., ``free running'', and thus phase noise accumulates over time. In contrast, if the LOs are locked using a phase-locked loop, the  phase noise is stationary, and can be modeled as, e.g., an \gls{iid} Gaussian process. In this case, a Kalman filter is not needed for phase tracking; the calibration is done using direct phase estimation following our scheduling scheme.}}
\revise{
\begin{align}
    \nut_{\ell,i} &= \nut_{\ell,i-1} + \deltat_{\ell,i}, \label{eq:PNt_Wiener} \\
    \nur_{\ell,i} &= \nur_{\ell,i-1} + \deltar_{\ell,i}, \label{eq:PNr_Wiener}
\end{align}
with $\deltat_{\ell,i}$ and $\deltar_{\ell,i}$} \gls{iid} as $\normal(0,\sigma_\nu^2)$ across {sample} time $i$. \revise{Here, we assume that the \glspl{AP} have identical \gls{LO} quality, represented by the variance
$\sigma_\nu^2$. According to~\cite[Eq.~(31)]{Piemontese2024}, this variance can be expressed as 
\begin{equation} \label{eq:PN_spectrum_level}
    \sigma_\nu^2 = 4 \pi^2 10^{10} S_{100} / f\sub{s}
\end{equation}
where 
$f\sub{s}$ is the signal bandwidth and $S_{100} (\dBcHz)$ is the phase noise spectrum level at $100\kHz$ offset from the center frequency. The phase noise spectrum level is typically given in the specification of commercial oscillators. Typical values are between $-120\dBcHz$ and $-80\dBcHz$ for realistic oscillators~\cite[\revise{Table 1}]{Piemontese2024}.}

Let the process $\{\alpha_n\}_{n = 1,2,\dots}$ collect the values of $\phi\of{2}_i - \phi\of{1}_i$ at instances $i_2$ of consecutive slots where phase estimation is performed. \revise{We next describe the Kalman filter for tracking this process.}
\begin{itemize}
    \item \textit{\revise{Process evolution model:}}
    The process evolution of $\alpha_n$ is
    \begin{equation}
        \alpha_{n} = \alpha_{n-1} + \zeta_{\revise{n-1}} \label{eq:kalman_process}
    \end{equation}
    where the noise $\{\zeta_{n}\}$ are \gls{iid} as $\normal(0,\sigma_\zeta^2)$ with $\sigma_\zeta^2$ representing the drift of \revise{$\nut_1$, $\nut_2$, $\nur_1$, and $\nur_2$ within the interval $[i_2-F\tc:i_2]$. As these phase noise terms drift independently with the same increment variance $\sigma_\nu^2$ over $F\tc$ samples of this interval, we have that $\sigma_\zeta^2 = \revise{4 \nsync\tau_c \sigma_\nu^2}$.} 
    
    \item \revise{\textit{Measurement model:}} Our observation of $\{\alpha_n\}$ is the sequence of $\bar \alpha_{i_2}$ in~\eqref{eq:2AP_phi_hat}. \revise{Two sources of measurement noise are present: i) the difference between $\alpha\of{1\to 2}_{i_2} -  \alpha\of{2\to 1}_{i_1}$ and $\phi_{i_2}\of{2} - \phi_{i_2}\of{1}$, caused by  the mismatch between the time index $[i_2]_{\floor{K/2}}$ (see~\eqref{eq:omega_ell} with $i = i_2$) and $i_1$ (see~\eqref{eq:tmp705}) of $\nur_1$ and $\nut_2$; ii) the error of estimating $\alpha\of{1\to 2}_{i_2} -  \alpha\of{2\to 1}_{i_1}$ by $\bar \alpha_{i_2}$, i.e., the total estimation error of $\alpha\of{2\to 1}_{i_1}$ and~$\alpha\of{1\to 2}_{i_2}$. The variance of the former noise is $2(i_1 - [i_2]_{\floor{K/2}}) \sigma_\nu^2 = 2(i_1 - {\floor{K/2}}) \sigma_\nu^2$. The variance of the latter noise is approximated as $\Pdl^{-1}\opnorm{\matG}^{-2}$.  Therefore, we use the following approximate measurement model for~$\alpha_n$:}
\begin{equation}
    \bar\alpha_{n} = \alpha_{n} + \xi_{n} + \mu_n \label{eq:kalman_observation}
\end{equation}
where $\xi_n \sim \normal(0,\sigma_\xi^2)$ with $\sigma_\xi^2 = 2(i_1-\floor{K/2})\sigma_\nu^2$, and $\mu_n \sim \normal\big(0,\Pdl^{-1}\opnorm{\matG}^{-2}\big)$. 
\item \textit{\revise{Kalman gain:}}
Notice that $\xi_n$ is contained in~$\zeta_{\revise{n-1}}$ because  $[[i_2]_{\floor{K/2}}:i_1] \subset [i_2-F\tc:i_2]$. Thus, the observation noise and process noise in~\eqref{eq:kalman_process} and~\eqref{eq:kalman_observation} are correlated. 
Specifically,
    $\Exop[\zeta_{\revise{n-1}}(\xi_n + \mu_n)] = \Exop[\xi_n^2] = \sigma_\xi^2$.
Therefore, we apply a generalized Kalman filter~\cite[Sec.~7.1]{Simon2006optimal}, \revise{described in Appendix~\ref{app:Kalman} to be self-contained,} that captures this correlation. 
In particular, the Kalman gain is computed~as
\begin{equation} \label{eq:2AP_Kalman_gain}
    \kappa_n = \frac{P_{n-1} + \sigma_\xi^2}{P_{n-1} + 3\sigma_\xi^2  + \Pdl^{-1}\opnorm{\matG}^{-2}}
\end{equation}
where $P_n$ is the current error variance. 

\item \textit{\revise{Update:}} The phase estimate is then updated as 
\begin{equation} \label{eq:2AP_update}
    \hat\alpha_{n} =  \hat\alpha_{n-1} + \kappa_n {\rm wrap}(\bar\alpha_n - \hat\alpha_{n-1} )
\end{equation}
where the wrapping function ${\rm wrap}(\alpha) = [(\alpha + \pi) \mod 2\pi] - \pi$ is used to resolve the circular nonlinearity~\cite{Markovic16}. The error variance is updated as 
\begin{equation} \label{eq:2AP_update_variance}
    P_{n}  = P_{n-1} - \kappa_n (P_{n-1} + \sigma_\xi^2) + \sigma_\zeta^2.
\end{equation}
\revise{The Kalman filter equations~\eqref{eq:2AP_Kalman_gain}--\eqref{eq:2AP_update_variance} follow straightforwardly from~\eqref{eq:Kalman_prior-update}--\eqref{eq:Kalman_posterior_update} in Appendix~\ref{app:Kalman}.}
\end{itemize}

The current filter output $\hat\alpha_{n}$ is used to estimate~$\phi\of{2}_i - \phi\of{1}_i$.

    %
    \revise{We can see that $\kappa_n$ becomes closer to $1$ if $\Pdl\opnorm{\matG}^2$ increases. That is, the Kalman filter output becomes closer to the direct measurement if the inter-\gls{AP} receive \gls{SNR} is improved, in which case the quality of the phase measurements is high. We shall verify this effect in Section~\ref{sec:result_2AP}.}
\section{Breaking TDD for Phase Calibration: \\The General Case}
\label{sec:general_case}
We now consider the general case with $L \ge 2$. We construct an undirected graph $\setG$ 
where nodes represent \glspl{AP}, \revise{and edges represent their connections.} 
We can adjust the topology of the graph $\setG$, such that $\setG$ is connected and has at least $L- 1$ edges. \revise{Typically, node~$i$ is connected to node~$j$ if the link between AP~$i$ and AP~$j$ is strong enough.} Bidirectional measurements are done only between \glspl{AP} connected by an edge in~$\setG$. 



\subsection{Measurement Model}
Recall that we need to estimate $\phi\of{\ell}_i$, $\ell \in [L]$, 
up to a common additive constant, at certain time $i$. 
Let $\vecphi_i = \tp{[\phi\of{1}_i \ \dots \ \phi\of{L}_i]}.$
Bidirectional phase measurements between AP~$\ell_1$ and AP~$\ell_2$ give an estimate of
\begin{equation}
    \alpha\of{\ell_1,\ell_2}_{i} =  \phi\of{\ell_2}_i - \phi\of{\ell_1}_i. 
    \label{eq:manyAP_measurements}
\end{equation}
We collect the values of $\alpha\of{\ell_1,\ell_2}_{i}$ across $M$ AP pairs $(\ell_1, \ell_2)$, $\ell_1 < \ell_2$, 
in a vector $\vecalpha_i$, and denote its estimate by $\hat \vecalpha_i$. 
Here, $M$ is the number of edges in $\setG$.
%
Similar to~\cite{Larsson24_synchrony}, we consider the following measurement model 
\begin{equation} \label{eq:angular_domain_model}
    \hat\vecalpha_i = \matB \vecphi_i + \jpg(\veczero, \matP_i) 
\end{equation}
where $\matP_i$ is a measurement error covariance matrix and 
$\matB$ is the $M \times L$ incidence matrix of $\setG$. 
For each measurement $m \in [M]$ corresponding to the AP pair $(\ell_1,\ell_2)$, the $(m,\ell_1)$th and $(m,\ell_2)$th 
elements of $\matB$ are given by $-1$ and $1$, respectively; 
all other elements in the $m$th row of $\matB$ are zero. Given $\hat\vecalpha_i$, we estimate~$\vecphi_i$, up to a common additive constant, as~\cite[Eq.~(22)]{Larsson24_synchrony}
\begin{equation} \label{eq:est_phase}
    \hat\vecphi_i = \matZ_i(\tp{\matZ}_i \tp{\matB} \matP_i^{-1} \matB \matZ_i)^{-1} \tp{\matZ}_i \tp{\matB} \matP_i^{-1} \hat\vecalpha_i
\end{equation}
where $\matZ_i$ contains the $L-1$ eigenvectors of $\tp{\matB}\matP_i^{-1} \matB$ that are orthogonal to the all-one vector. 
Note that this is a least-squares estimate only if the elements of $\vecalpha_i$ are small, so that the implicit mod $2\pi$ operator in~\eqref{eq:angular_domain_model} becomes the identity mapping, as assumed in~\cite{Larsson24_synchrony}. To reduce the impact of circular nonlinearity, we unwrap $\vecalpha_i$ before computing $\hat\vecphi_i$. Specifically, whenever the jump of each element from $\hat\vecalpha_{i-1}$ to $\hat\vecalpha_i$  
exceeds $\pi$ (in absolute value), we 
add multiples of $\pm2\pi$ until the jump is less than $\pi$.


In the remainder of the section, we focus on the acquisition of $(\hat\vecalpha_i,\matP_i)$ using a broken TDD flow. 

\subsection{Phase Measurement}
We divide time into frames of $\nsync$ slots, and select $\nmeas$ slots of each frame as the \emph{measurement slots}. The pattern of the measurement slots is a design choice. We let a selected set of \glspl{AP} shift the timing of the uplink/downlink TDD switching point 
to enable phase shift estimation, such that a new estimate of every component of $\vecalpha$ is obtained within the $\nmeas$ measurement slots. The frame structure is illustrated in Fig.~\ref{fig:frame}.
A straightforward extension from the two-AP case is to dedicate one measurement slot to each AP pair. 
However, the required number of slots, $\nmeas$, is as large as $M$, which scales quadratically with $L$ if $\setG$ is complete. We next propose a scheme to reduce~$\nmeas$. 

\begin{figure}
    \centering
    \scalebox{.9}{
    \begin{tikzpicture}[>=Stealth]

    \tikzset{
      slot/.style={draw, minimum width=1.4cm, minimum height=.4cm, align=center},
      meas/.style={draw, rounded corners, inner xsep=6pt, inner ysep=4pt}
    }
    
    \node[slot, pattern=north east lines] (r1) {};
    \node[slot, right=0cm of r1]        (r2) {};
    \node[slot, right=0cm of r2, pattern=north east lines] (r3) {};
    \node[slot, right=0cm of r3]        (r4) {$\cdots$};
    \node[slot, right=0cm of r4, pattern=north east lines] (r5) {};
    \node[slot, right=0cm of r5]        (r6) {};
    
    \node[below=1pt of r1] {slot $1$};
    \node[below=1pt of r6] {slot $F$};
    
    \node[above=.3cm of r3] (measure) {$\nmeas$ measurement slots};
    
    \draw[->] (r1.north) to[out=70,in=-180] (measure.west);
    \draw[->] (r3.north) to[out=90,in=-90] (measure.south);
    \draw[->] (r5.north) to[out=90,in=0]  (measure.east);
    
    \end{tikzpicture}
    }
        


    \caption{Illustration of a frame of $\nsync$ slots over which the TDD flow is broken for some APs within $\nmeas$ selected slots, called measurement slots, to enable phase measurements.}
    \label{fig:frame}
\end{figure}

\subsubsection{Complete Graph Topology}
Consider first the case where all \glspl{AP} are connected to each other.
We construct a broken TDD flow with $\nmeas = L - 1$ measurement slots as follows. In each measurement slot, we let an \gls{AP}, called the \textit{master \gls{AP}}, shift its periods in the same way as \gls{AP}~2 in Fig.~\ref{fig:TDD_operation}(b), while all other \glspl{AP} do not change their slot structure. Within the slot, the master \gls{AP} sends a calibration signal to all other APs at time~$i_1$, and receives calibration signal from another AP at time~$i_2$. By letting $L-1$ APs take turns to play the role of the master AP, and letting the remaining AP send calibration signal to the current master AP in each slot, we ensure that bidirectional phase measurements are done for all AP pairs. A detailed signal model of this scheme is given below.

Let us index the measurement slots by $1,\dots,
L-1$. 
In the $j$th measurement slot, we let \gls{AP}~$j+1$ be the master AP. At time $i_1$ of this slot, \gls{AP}~$j+1$ 
transmits a calibration signal 
\begin{equation}
    \vecx\supp{sync}_{j+1} = 
     \sum_{\ell \in \setN_{j+1}} \sqrt{\rho_{j+1, \ell}} \vecu_{j+1,\ell} \label{eq:sync_signal_masterAP}
\end{equation}
where $\setN_{j+1}$ denotes the set of neighbors of AP~$j+1$ in graph~$\setG$, $\vecu_{j+1,\ell}$ is the leading right-singular vector of the matrix $\matG_{\ell,j+1}$, and $\{\rho_{j+1, \ell}\}_\ell$ are power allocation terms such that $\sum_{\ell \in \setN_{j+1}} \rho_{j+1, \ell} = \Pdl$. A heuristic power allocation scheme, inspired by fractional power allocation proposed in~\cite{Interdonato19}, is 
\begin{equation}
    \rho_{j+1, \ell} = \Pdl \frac{\|\matG_{j+1, \ell}\|^{-1}}{\sum_{\ell' \in \setN_{j+1}} \|\matG_{j+1, \ell'}\|^{-1} }, \quad \ell \in \setN_{j+1}. \label{eq:FPA_sync}
\end{equation}
This way, more power is allocated to weaker inter-AP links, so that the phase measurement error is balanced between the links.
The received signal at \gls{AP}~$\ell$, $\ell \in \setN_{j+1}$, is given by
\begin{equation}
    \vecy_{\ell,i_1}\supp{sync} = 
    \exp\big(\jmath \alpha\of{j+1 \to \ell}_{i_1}\big) \matG_{\ell,j+1} 
    \vecx\supp{sync}_{j+1} + \vecz_{\ell,i_1}\supp{sync}, \label{eq:manyAP_sync_l}
\end{equation}
where $\alpha\of{j+1 \to \ell}_{i_1} =  \revise{-\nut_{j+1,i_1} + \nur_{\ell,i_1}}$. 
\gls{AP}~$\ell$ estimates $\alpha\of{j+1 \to \ell}_{i_1}$ by
$\bar \alpha\of{j+1 \to \ell}_{i_1} = \angle \herm{\vecu}_{j+1,\ell} \herm{\matG}_{\ell,j+1} \vecy_{\ell,i_1}\supp{sync}$ 
and the \gls{MSE} is approximated as $\dfrac{\vecnorm{\matG_{\ell,j+1} \vecu_{j+1,\ell}}^2}{2|\herm{\vecu}_{j+1,\ell} \herm{\matG}_{\ell,j+1} \matG_{\ell,j+1} \vecx\supp{sync}_{j+1}|^2}$.

At time $i_2$ of the slot, \gls{AP}~$1$ transmits the calibration signal
    $\vecx\supp{sync}_{1} \!=\! 
    \sqrt{\Pdl} \vecu_{1,j+1}$ where $\vecu_{1,j+1}$ is the leading right-singular vector of $\matG_{j+1,1}$.
\gls{AP}~$j+1$ is in uplink mode and receives
\begin{equation}
    \vecy_{j+1,i_2}\supp{sync} = \exp\big(\jmath \alpha\of{1 \to j+1}_{i_2}\big) \matG_{j+1,1} 
    \vecx\supp{sync}_{1} + \vecz_{j+1,i_2}\supp{sync}
\end{equation}
with $\alpha\of{1 \to j+1}_{i_2} =  \revise{-\nut_{1,i_2} + \nur_{j+1,i_2}}$, 
and estimates $\alpha\of{1 \to j+1}_{i_2}$ by
$\bar \alpha\of{1 \to j+1}_{i_2} = \angle \herm{\vecu}_{1,j+1} \herm{\matG}_{j+1,1} \vecy_{j+1,i_2}\supp{sync}$. We approximate the \gls{MSE} of this estimate as $$\dfrac{\vecnorm{\matG_{j+1,1} \vecu_{1,j+1}}^2}{2|\herm{\vecu}_{1,j+1} \herm{\matG}_{j+1,1} \matG_{j+1,1} \vecx\supp{sync}_{1}|^2} = (2\Pdl \opnorm{\matG_{j+1,1}}^2)^{-1}.$$


After receiving the phase estimates from \gls{AP}~$\ell_1$ and \gls{AP}~$\ell_2$, the \gls{CPU} takes
\begin{equation} \label{eq:measure_l1_l2}
    \bar \alpha\of{\ell_1,\ell_2}_{\max\{i_{\ell_1}, i_{\ell_2}\}} = \bar \alpha_{i_{\ell_2}}\of{\ell_1 \to \ell_2} - \bar \alpha_{i_{\ell_1}}\of{\ell_2 \to \ell_1}    
\end{equation}
as an estimate of $\alpha\of{\ell_1,\ell_2}_{\max\{i_{\ell_1}, i_{\ell_2}\}}$, where $i_{\ell_1}$ and $i_{\ell_2}$ are indices of the samples where calibration signal is sent from AP~$\ell_2$ to AP~$\ell_1$ and from AP~$\ell_1$ to AP~$\ell_2$, respectively. 

\subsubsection{General Connected Graph Topology}

We now consider the general case where some APs may not be connected in~$\setG$.
The APs that do not share neighbors in $\setG$ can transmit simultaneously without causing significant interference to their unintended receivers. For example, for the configuration shown in Fig.~\ref{fig:graph_manyAP}, AP~$4$ and AP~$5$ can transmit at the same time. Therefore, we let them break their TDD flow in the same slot and send calibration signal to their respective neighbors.
Following this principle, we propose the following measurement scheme.
We first perform a distance-$2$ coloring of $\setG$ (such that two nodes that are adjacent or share a neighbor do not share color), resulting in $n\sub{c}$ colors. We group the \glspl{AP} in terms of their color. In each of $n\sub{c}-1$ slots, we select one of the last $n\sub{c}-1$ groups and let the \glspl{AP} therein be master APs simultaneously, i.e., they shift their TDD flow and transmit calibration signal at time $i_1$ of the slot. We also let each \gls{AP} in the first group transmit a calibration signal to the current master \gls{AP} in its neighbor at time $i_2$ of the slot. (Note that at most one master AP is in the neighborhood of a given node.) This way, the number of required measurement slots is $\nmeas = n\sub{c}-1$, whose minimum value is $\chi_2(\setG) - 1$,
where $\chi_2(\setG)$ is the distance-$2$ chromatic number of $\setG$.
The complete-graph case is a special case of this scheme where $\chi_2(\setG) = L$.


In Fig.~\ref{fig:slots_manyAP}, we demonstrate the measurement slots for the example in Fig.~\ref{fig:graph_manyAP}. The graph has distance-$2$ chromatic number $\chi_2(\setG) = 4$, and it takes $\chi_2(\setG) - 1 = 3$ slots to obtain a new measure of $\vecalpha$. APs~$4$ and $5$ simultaneously serve as the master APs in the first slot, while AP~$2$ and AP~$3$ take this role in slot~$2$ and slot~$3$, respectively. 
\begin{figure}
    \centering
    \captionsetup[subfigure]{oneside,margin={0cm,0.25cm}}
    \subcaptionbox{The graph and its $2$-distance coloring. AP~$4$ and AP~$5$ are given the same color, i.e., they can transmit simultaneously without causing interference to each other's neighbors. \label{fig:graph_manyAP}}{
        \scalebox{.9}{\begin{tikzpicture}
    
        \tikzstyle{AP} = [circle, draw, thick, minimum size=.5cm, node distance=2cm]
        
        \node[AP, fill = yellow!60] (n1) at (0,0) {1};
        \node[AP, fill=red!30] (n2) at (1.4,0) {2};
        \node[AP, fill=green!50] (n3) at (1.4,-1.4) {3};
        \node[AP, fill = blue!30] (n4) at (0,-1.4) {4};
        \node[AP, fill= blue!30] (n5) at (2.8,0) {5};

        \draw (n1) -- (n2);
        \draw (n2) -- (n3);
        \draw (n3) -- (n4);
        \draw (n4) -- (n1);
        \draw (n5) -- (n2);
        \draw (n1) -- (n3);
        
        \end{tikzpicture}}
    }
    \captionsetup[subfigure]{oneside,margin={0cm,0.25cm}}
    \subcaptionbox{Three measurement slots over which bidirectional measurements between all connected AP pairs are done. Shaded boxes represent slots where the indicated \gls{AP} shifts its TDD periods to send and receive calibration signals at time $i_1$ and $i_2$, respectively. Arrows indicates the direction of calibration signals. \label{fig:slots_manyAP}}{
        \hspace{-.5cm}
        ~\scalebox{.9}{
        \begin{tikzpicture}
            \def\cellwidth{1.5cm}
            \def\cellheight{.4cm}

            \foreach \j in {1,2,3,4,5} {
            \foreach \i in {1,2,3} {
                \draw (\i*\cellwidth, -\j*1.5*\cellheight) rectangle ++(\cellwidth, \cellheight);
            }
            \node at (1cm,-\j*1.5*\cellheight+.5*\cellheight) {AP $\j$};
            }

            \foreach \i/\j in {1/4,1/5,2/2,3/3} {
                \draw[fill=gray!50] (\i*\cellwidth, -\j*1.5*\cellheight) rectangle ++(\cellwidth, \cellheight);
            }

            \foreach \i/\ja/\jb in {1/4/1,1/4/3,1/5/2,2/2/1,2/2/3,2/2/5,3/3/1,3/3/2,3/3/4,1/1/4,2/1/2,3/1/3} {
                \ifnum \ja = 1
                    \def\ll{.6}
                    \def\benda{right}
                    \def\bendb{left}
                \else
                    \def\ll{.4}
                    \def\benda{left}
                    \def\bendb{right}
                \fi
                
                \ifnum \ja > \jb
                    \draw[bend \benda=30,-latex] (\i*\cellwidth + \ll*\cellwidth, -\ja*1.5*\cellheight+.5*\cellheight) to (\i*\cellwidth + \ll*\cellwidth, -\jb*1.5*\cellheight+.5*\cellheight);
                \else
                    \draw[bend \bendb=30,-latex] (\i*\cellwidth + \ll*\cellwidth, -\ja*1.5*\cellheight+.5*\cellheight) to (\i*\cellwidth + \ll*\cellwidth, -\jb*1.5*\cellheight+.5*\cellheight);
                \fi
            }


        \end{tikzpicture} ~~
        }
    }
    \caption{An example of five \glspl{AP} and their connections. It takes $\chi_2(\setG) = 4$ colors to color the APs, and thus $\nmeas = \chi_2(\setG) - 1 = 3$ slots.}
    \label{fig:example_manyAP}
\end{figure}

We omit the signal model for the measurements, as this can be written similarly to the complete-graph case.  

\subsection{Kalman Filter}
Here, we consider again the Wiener phase noise model \revise{in~\eqref{eq:PNt_Wiener} and~\eqref{eq:PNr_Wiener}.} 
Let the process $\{\vecalpha_n\}$ collect the true values of $\vecalpha$ at time $i_2$ of consecutive measurement slots. \revise{We next describe the Kalman filter to track this process.} 

\begin{itemize}
    \item \textit{\revise{Process evolution model:}} Due to phase drift, the process evolution of~$\vecalpha_n$ is
    \begin{equation}
        \vecalpha_{n} = \vecalpha_{n-1} + \veczeta_{\revise{n-1}} \label{eq:kalman_process_manyAP}
    \end{equation}
    where the process noise terms $\veczeta_{\revise{n-1}}$ \revise{represents the drift of $\nut_{\ell_1}$,  $\nut_{\ell_2}$, $\nur_{\ell_1}$, and $\nur_{\ell_2}$ between consecutive measurement slots. It} is distributed as $\normal(\veczero,\matSigma_{\revise{n-1}}\of{\veczeta})$ with $\matSigma_{n}\of{\veczeta}$ computed in Appendix~\ref{sec:compute_cov_zeta}.

    \item \textit{\revise{Measurement model:}} \revise{In each measurement slot, an} observation of $\vecalpha_n$ is obtained from the phase measurements in the latest $\nmeas$ measurement slots. This observation is partial, as a new measurement is available for only the APs that transmit or receive the calibration signal in the current measurement slot. For the example in Fig.~\ref{fig:slots_manyAP}, at the end of the first, second, and third measurement slots, a new observation of $\{\alpha\of{1,4},\alpha\of{2,5},\alpha\of{3,4}\}$, $\{\alpha\of{1,2},\alpha\of{2,3}, \alpha\of{2,5}\}$ and $\{\alpha\of{1,3},\alpha\of{2,3},\alpha\of{3,4}\}$, respectively, are obtained.
    The observation model can be approximated as
    \begin{equation}
        \bar\vecalpha_{n} = \matA_n \vecalpha_{n} + \vecxi_{n} + \vecmu_n  \label{eq:kalman_observation_manyAP}
    \end{equation}
    where $\matA_n$ is the measurement matrix (containing rows of $\matidentity_M$) that picks AP pairs $(\ell_1,\ell_2)$ having a new measurement of $\alpha\of{\ell_1,\ell_2}$. \revise{Furthermore,} $\vecxi_n \sim \normal(\veczero,\matSigma_n\of{\vecxi})$ is the noise due to the offset between time indices of the terms in~\eqref{eq:manyAP_measurements} and the time when they are measured. \revise{We compute $\matSigma_n\of{\vecxi}$ in Appendix~\ref{sec:compute_cov_xi}. Finally,} 
    $\vecmu_n \sim \normal\big(\veczero, \matSigma_n\of{\vecmu}\big)$ 
    represents the total measurement error of $\alpha\of{\ell_1\to \ell_2}_{i_{\ell_2}}$ and~$\alpha\of{\ell_2\to \ell_1}_{i_{\ell_1}}$. Here,  $\matSigma_n\of{\vecmu}$ is a diagonal matrix with diagonal element corresponding to $(\ell_1, \ell_2)$ given by $\sigma^2_{\ell_1 \to \ell_2} + \sigma^2_{\ell_2 \to \ell_1}$ with 
        $\sigma^2_{i \to j} = \dfrac{\vecnorm{\matG_{j,i} \vecu_{i,j}}^2}{2|\herm{\vecu}_{i,j} \herm{\matG}_{j,i} \matG_{j,i} \vecx_{i}|^2}$.

    \item \textit{\revise{Kalman gain:}} The process noise in~\eqref{eq:kalman_process_manyAP} and observation  noise in~\eqref{eq:kalman_observation_manyAP} are correlated, \revise{as they are partially caused by phase drifts over overlapping intervals.} 
    The correlation matrix is given by
        $\Exop[\veczeta_{\revise{n-1}}\tp{(\vecxi_n + \vecmu_n)}] = \Exop[\veczeta_{\revise{n-1}} \tp{\vecxi}_n] = \matSigma_{\revise{n-1}}\of{\veczeta\vecxi}$,
    computed in Appendix~\ref{sec:compute_corr}.
    Therefore, as in the two-AP case, we apply the generalized Kalman filter~\cite[Sec.~7.1]{Simon2006optimal} \revise{presented in Appendix~\ref{app:Kalman}}. 
    In particular, given the previous posterior error covariance matrix $\matP^+_{n-1}$, the CPU computes the current prior error covariance matrix as
        $\matP_n^- = \matP^+_{n-1} + \matSigma_n\of{\veczeta}$
    and the Kalman gain as
    \begin{multline}
        \!\!\!\!\matK_n = (\matP^-_{n} \tp{\matA}_n + \matSigma_n\of{\veczeta\vecxi}) \\ \cdot (\matA_n \matP^-_{n} \tp{\matA}_n + \matA_n\matSigma_n\of{\veczeta\vecxi} + \tp{({\matSigma}_n\of{\veczeta\vecxi})} \tp{\matA}_n + \matSigma_n\of{\vecxi} + \matSigma_n\of{\vecmu})^{-1}. \label{eq:Kalman_gain}
    \end{multline}

    \item \textit{\revise{Update:}}
    The phase estimate is then updated as 
    \begin{equation}
        \hat\vecalpha_{n} = \hat\vecalpha_{n-1} + \matK_n {\rm wrap}(\bar\vecalpha_n - \matA_n \hat\vecalpha_{n-1}) 
    \end{equation}
    where the wrapping function is applied element-wise. The posterior error covariance is updated as 
    \begin{equation} \label{eq:update_variance}
        \matP^+_{n}  = \matP^-_{n} - \matK_n (\matA_n \matP^-_{n} + \tp{(\matSigma_n\of{ \veczeta\vecxi})}).
    \end{equation}
    \revise{As in the two-\gls{AP} case, the Kalman filter equations~\eqref{eq:Kalman_gain}--\eqref{eq:update_variance} follow straightforwardly from~\eqref{eq:Kalman_prior-update}--\eqref{eq:Kalman_posterior_update} in Appendix~\ref{app:Kalman}.}
\end{itemize}

The current filter output $(\hat\vecalpha_{n}, \matP^+_n)$ is used to estimate $\vecphi_i$ as in~\eqref{eq:est_phase}. 

\revise{
\subsection{Complexity Analysis} \label{sec:complexity}
    We analyze the complexity of the operations done at the central server. 
    The distance-2 coloring problem is in general NP-hard, but one can use an adaptation of the greedy Welsh–Powell algorithm~\cite{Welsh67}, which has complexity order $O(L \Delta^2 + L \log L)$ where $\Delta$ is the maximum vertex degree of graph $\setG$. 
    Given the graph $\setG$ and the frame structure, the matrices $\matSigma_n\of{\vecxi}$, $\matSigma_n\of{\vecmu}$, $\matSigma_n\of{\veczeta}$, and $\matSigma_n\of{\veczeta\vecxi}$ can be computed with complexity order $O(M^2)$. The complexity of each Kalman filter iteration is $O(M^3)$, dominated by the computation of the Kalman gain in~\eqref{eq:Kalman_gain}. Finally, the computation of $\hat\vecphi_i$ in~\eqref{eq:est_phase} also has complexity order $O(M^3)$. Recall that $M \ge L-1$. We conclude that the overall complexity order is $O(M^3)$. Therefore, limiting $M$, e.g., in the same order as $L$, is beneficial in terms of complexity, and thus the scalability of the phase tracking process as $L$ grows.
}

\section{Achievable Spectral Efficiency} \label{sec:rate}

We derive the achievable \gls{SE} of downlink beamforming after phase calibration. Following~\cite{marzetta2016fundamentals}, we write the received signals as a deterministic gain times the signal of interest plus uncorrelated noise. The noise comprises beamforming gain uncertainty, multiuser interference, and \gls{AWGN}. We then use the ``use and forget'' trick~\cite[Sec.~2.3.2]{marzetta2016fundamentals} to lower-bound the ergodic capacity,  inherently assuming that every codeword spans many realizations of all randomness, including the phase calibration errors. 

Let $\psi_{k,i}$ be the phase compensation term of \gls{UE}~$k$ at sample~$i$. That is, $-\psi_{k,i}$ is \gls{UE}~$k$'s current estimate of the phase shift $c$ (recall~\eqref{eq:manyAP_theta}). We  express the effective received signal at time~$i$ as
\begin{align}
    e^{\jmath {\psi_{k,i}}} y_{k,i} &= \DS_{k,i} s_{k,i} + \BU_{k,i} s_{k,i} + \UI_{k,i} + e^{\jmath {\psi_{k,i}}} z_{k,i}. \label{eq:rx_conventionalTDD}
\end{align}
where for conjugate beamforming,
\begin{align}
    \DS_{k,i} &= \sqrt{\Pdl}\Exop\left[\sum_{\ell = 1}^L a_{\ell,i} \sqrt{\frac{\eta_{k,\ell}}{N \gamma_{k,\ell}}} \Delta_{k,\ell,i} \tp{\vecq}_{k,\ell,i} \hat\vecq_{k,\ell,i}^*\right], \label{eq:tmp410} \\ 
    \!\!\!\BU_{k,i} &= \!\sqrt{\Pdl\!} \sum_{\ell = 1}^L \!a_{\ell,i} \sqrt{\!\frac{\eta_{k,\ell}}{N \gamma_{k,\ell}}} \Delta_{k,\ell,i} \tp{\vecq}_{k,\ell,i} \hat\vecq_{k,\ell,i}^* \!- \DS_{k,i}, \! \label{eq:conj_BU}\\ 
    \UI_{k,i} &= \!\!\sum_{k'=1, k'\ne k}^K   \underbrace{\!\!\Bigg(\!\sqrt{\Pdl}\sum_{\ell = 1}^L a_{\ell,i}  \sqrt{\frac{\eta_{k',\ell}}{N \gamma_{k',\ell}}}  \Delta_{k,\ell,i} \tp{\vecq}_{k,\ell,i} \hat\vecq_{k',\ell,i}^* \!\Bigg)\!}_{= \UI_{k,k',i}} \notag \\
    &\quad \cdot s_{k',i}, 
\end{align}
and for \gls{ZF} beamforming,
\begin{align}
    \DS_{k,i} &=  \sqrt{(N-K)\Pdl} \sum_{\ell = 1}^L a_{\ell,i} \sqrt{\eta_{k,\ell} \gamma_{k,\ell}} \Exop[\Delta_{k,\ell,i}],  \\ 
    \!\!\!\BU_{k,i} &=  \sqrt{(N\!-\!K)\Pdl\!} \sum_{\ell = 1}^L a_{\ell,i} \sqrt{\eta_{k,\ell}\gamma_{k,\ell}} \Delta_{k,\ell,i} \!- \DS_{k,i},  \label{eq:ZF_BU}\\ 
    \UI_{k,i} &= \sqrt{\Pdl} \sum_{\ell = 1}^L   a_{\ell,i} \Delta_{k,\ell,i} \tp{\tilde \vecq}_{k,\ell,i} \matW_{\ell,i}  \matD_{\veceta_\ell}^{1/2} \vecs_{i}. \label{eq:tmp418}
\end{align}
Here, the term
\begin{equation}
    \Delta_{k,\ell,i} = \exp[\jmath (-\revise{\nut_{\ell,i}} - \revise{\nur_{\ell,[i]_k}} + {\theta_{\ell,i}} + {\psi_{k,i}})]. \label{eq:Delta}
\end{equation}
represents residual multiplicative phase noise after compensation.
The terms $\DS_{k,i}$, $\BU_{k,i}$, and $\UI_{k,i}$ represent the strength of the desired signal (DS), the beamforming gain uncertainty (BU), and the (residual) interference, respectively.

\subsection{Achievable Rate}
Let $\setJ = \{j_1, j_2, \dots\}$ be a sequence of samples across slots such that $\{\Delta_{k,\ell,i}\}_{i \in \setJ}$ and $\{a_{\ell,i}\}_{i \in \setJ}$ are ergodic processes. For the process $\{a_{\ell,i}\}_{i \in \setJ}$, it suffices to have that $a_{\ell,i}$ remains constant (either $0$ or $1$) for $i \in \setJ$.
We derive the achievable rate when channel coding is performed across $i \in \setJ$. 

For $i\in \setJ$, we can easily verify that the effective noise, containing the last three terms in~\eqref{eq:rx_conventionalTDD}, is uncorrelated with the desired signal for both beamforming schemes. Therefore, using the {``use and forget'' trick~\cite[Sec.~2.3.2]{marzetta2016fundamentals}} 
and the fact that uncorrelated Gaussian noise represents the worst case, we obtain an achievable rate for coding over the sample sequence $\setJ$, represented by time index~$i$, as 
    \begin{align}
        R_{k,i} = \log_2\left(1+\frac{\abs{\DS_{k,i}}^2}{\Exop\big[\abs{\BU_{k,i}}^2\big] + 
        \Exop\big[\abs{\UI_{k,i}}^2\big] + 1} \right). \label{eq:rate_i}
    \end{align}
In the next theorem, we provide an expression for this achievable rate where the expectations are computed in closed form except for the randomness of $\Delta_{k,\ell,i}$. 

\begin{thm}
\label{th:rate_i}
    An achievable rate of the downlink transmission from the \glspl{AP} to user~$k$ when channel coding is applied across the sequence of samples $\setJ$, represented by time index~$i$, is given by~\eqref{eq:rate_i_conj} and~\eqref{eq:rate_i_ZF}, shown at the top of the next page, for conjugate beamforming and \gls{ZF} beamforming, respectively. 
    \begin{figure*}[t]
        \begin{align}
            R_{k,i} &= 
            \log_2\Bigg(1+\frac{N \Pdl\abs{ \sum_{\ell = 1}^L a_{\ell,i} \sqrt{\eta_{k,\ell} \gamma_{k,\ell}} \Exop\left[\Delta_{k,\ell,i}\right]}^2}{N {\Pdl}\sum_{\ell = 1}^L a_{\ell,i} \eta_{k,\ell} \gamma_{k,\ell} \big(1 - \left|\Exop\left[\Delta_{k,\ell,i}\right]\right|^2\big) +  \Pdl  \sum_{\ell = 1}^L a_{\ell,i} \beta_{k,\ell} \sum_{k' = 1}^K \eta_{k',\ell}   +  1} \Bigg), \label{eq:rate_i_conj} \\ 
            R_{k,i} &= 
            \log_2\Bigg(1+\frac{(N-K)\Pdl\abs{ \sum_{\ell = 1}^L a_{\ell,i} \sqrt{\eta_{k,\ell} \gamma_{k,\ell}}  \Exop\left[\Delta_{k,\ell,i}\right]}^2}{(N-K){\Pdl}\Varop\!\left[\sum_{\ell = 1}^L a_{\ell,i} \sqrt{\eta_{k,\ell} \gamma_{k,\ell}} \Delta_{k,\ell,i}\right] +  \Pdl \sum_{\ell = 1}^L   a_{\ell,i} (\beta_{k,\ell} - \gamma_{k,\ell}) \sum_{k' = 1}^K \eta_{k',\ell}  +  1} \Bigg), \label{eq:rate_i_ZF}
        \end{align}
        \hrule
    \end{figure*}
    
\end{thm}
\begin{proof}
    See Appendix~\ref{proof:rate_i}.
\end{proof}


    Comparing the rate formulas in~\eqref{eq:rate_i_conj} and~\eqref{eq:rate_i_ZF} with the no-phase-noise counterparts in~\cite[Eq.~(24)]{Ngo17_CF} and~\cite[Eq.~(21)]{Interdonato19_ZF}, 
    respectively, we 
        see that phase noise leads to a down-scaling of the DS strength (the numerators in~\eqref{eq:rate_i_conj} and~\eqref{eq:rate_i_ZF})
        by  a factor $\Exop[\Delta_{k,\ell,i}]$, and a new BU term  (the first term in the denominators in~\eqref{eq:rate_i_conj} and~\eqref{eq:rate_i_ZF}). 
 %
        For \gls{ZF} beamforming, the new BU term 
        results only from the variation of the collective residual phase noise around its mean (see~\eqref{eq:ZF_BU}), whereas this term also involves the propagation channel in the case of conjugate beamforming (see~\eqref{eq:conj_BU}). 
        Furthermore, phase noise does not affect the strength of inter-user interference (the second term in the denominators in~\eqref{eq:rate_i_conj} and~\eqref{eq:rate_i_ZF}).
        %
        
        We have that $|\Exop[\Delta_{\ell,i,k}]| \le 1$, with equality achieved when phase noise is not present or is perfectly compensated for. 
        Imperfect compensation of phase noise makes $|\Exop[\Delta_{\ell,i,k}]|$ strictly smaller than $1$, and thus decreases the rate. Recall that $\theta_{\ell,i}$ is reset in each measurement slot. 
        The process $\{\Delta_{k,\ell,i}\}_{i \in \setJ}$ contains three main components: 1) the phase drift during the offset between time $i$ and the latest time when $\theta_{\ell,i}$ and $\psi_{k,i}$ are reset, 2) the error of estimating $\hat \vecalpha$ and $\psi_{k,i}$ from bidirectional and downlink measurements, respectively, and 3) the error of solving for $\vecphi_i$ from $\hat \vecalpha_i$. The first component increases with the phase drift variance $\sigma_\nu^2$. The second component depends on the received \gls{SNR} of the phase \revise{calibration} signals. The third component depends on the topology of~$\setG$ as analyzed in~\cite{Larsson24_synchrony}. In particular, when $L$ increases, the error variance of estimating $\vecphi_i$ from $\hat \vecalpha_i$ can grow unbounded for some graph topologies, such as the line topology.


\subsection{Achievable Spectral Efficiency}
We divide the time horizon into frames of $\nsync$ slots in the same manner described in Fig.~\ref{fig:frame}. Let $i_{f,1}, \dots, i_{f,\nsync \tc}$ be the indices of the samples in frame~$f$. Each cross-frame sequence $\setI_n = \{i_{1,n}, i_{2,n}, \dots\}$, $n\in [\nsync \tc]$, satisfy the assumptions of the sequence $\setJ$ described in the previous subsection. Specifically, for each $n\in[\nsync\tc]$, the three mentioned components of process $\{\Delta_{k,\ell,i}\}_{i \in \setI_n}$ are ergodic.
We apply multiple channel codes, one for each of the $\nsync \tc$ sequences $\setI_n$, $n\in[\nsync \tc]$, and leverage Theorem~\ref{th:rate_i} to obtain the following achievable \gls{SE}. 
\begin{cor} \label{cor:SE}
    An achievable downlink \gls{SE} of \gls{UE}~$k$ is
    \begin{equation}
        \SEdl_k = \frac{1}{\nsync \tc} \sum_{n=1}^{\nsync \tc} R_{k,n} \quad \text{bit/s/Hz}
    \end{equation}
    where 
    $R_{k,n}$ is the achievable rate for the sequence $\setI_n$, computed as in~\eqref{eq:rate_i_conj} for conjugate beamforming and in~\eqref{eq:rate_i_ZF} for \gls{ZF} beamforming. 
\end{cor}

Compared to the conventional TDD flow, the shifted TDD structure incurs a pre-log SE loss. Specifically, in a frame, $\nmeas$ samples of the downlink periods are used for calibration measurements instead of data transmission. Furthermore, in each measurement slot, the master APs (with shifted TDD structure) do not transmit data during $\tg$ samples where their guard period overlaps with the downlink period of the slave APs. However, phase calibration increases the magnitude of $\Delta_{k,\ell,i}$, thus improving the rate for all other downlink samples. 
\section{Numerical Experiments}
\label{sec:result}
We numerically evaluate the achievable \gls{SE} of the proposed broken \gls{TDD} flow. The common parameters chosen in all experiments are summarized in Table~\ref{tab:parameters}. Therein, we show the values of maximum radiated power $\bar{\rho}\sub{UE}$ per \gls{UE} and $\bar{\rho}\sub{AP}$ per \gls{AP}. The corresponding normalized transmit powers $\Ppl$ and $\Pdl$ are obtained by dividing these powers by the noise power, $-94\dBm$, \revise{which consists of thermal noise for the considered bandwidth with a noise figure of $7\dB$. The calibration signals are transmitted with the same power as the downlink data, unless otherwise stated.} 
We consider a square simulation area of $500 \times 500 \m^2$, \revise{and wrap it around with eight identical neighbor areas.} 
The UEs are uniformly distributed at random within the area. As in~\cite{Demir21}, we adopt the 3GPP Urban Microcell propagation model in~\cite[Tab. B.1.2.1-1]{3gpp.36.814} with carrier frequency $f\sub{c} = 2 \GHz$, for which
\begin{equation}
    \beta_{k,\ell} ~[\!\dB] = -30.5 - 36.7 \log_{10} d_{k,\ell} + S_{k,\ell} \label{eq:propagation_model}
\end{equation}
where $d_{k,\ell}$ is the distance, in meters, between UE~$k$ and AP~$\ell$ and $S_{k,\ell} \sim \normal(0,4^2)$ is the shadow fading. We consider correlated fading from an AP to different UEs with $\Exop[S_{k,\ell} S_{i,j}]$ given by $4^2 2^{-d\supp{UE}_{k,i}/9 \m}$ if $\ell = j$ and by $0$ otherwise, where $d\supp{UE}_{k,i}$ is the distance between UE~$k$ and UE~$i$~\cite[Tab. B.1.2.2.1-4]{3gpp.36.814}. The inter-AP channels $\matG_{i,j}$ have \gls{iid} $\jpg(0,\beta_{i,j}\supp{AP})$ entries, where $\beta_{i,j}\supp{AP}$ follows the same model as~\eqref{eq:propagation_model}. The correlated shadowing for the inter-AP links is negligible as the APs are at least $50\m$ apart. We adopt the fractional power allocation scheme~\cite{Interdonato19} where
\begin{equation}
    \eta_{k,\ell} = \frac{\beta_{k,\ell}^{1/2}}{ \sum_{k'=1}^K \beta_{k',\ell}^{1/2}}, \quad k\in [K], \ell \in [L]. 
\end{equation}
\revise{The considered setting and parameters resemble widely-considered scenarios in the literature, e.g.,~\cite{Ngo17_CF,Demir21,Interdonato18}.}

\begin{table}[t!] \small
    \centering
    \caption{\small \sc System parameters}
    \label{tab:parameters}
    \begin{tabular}{|c|c|}
         \hline Parameter & Value \\ \hline\hline
         Simulation area & $500 \times 500 \m^2$ \\ \hline 
          Number of \glspl{UE}, $K$ & \revise{from $5$ to $20$} \\\hline
         Carrier frequency, $f\sub{c}$ & $2\GHz$ \\\hline
         Signal bandwidth, $f\sub{s}$ & $20\MHz$ \\\hline
         Slot length, $\tc$ & $100$ samples \\\hline
         Guard period, $\tg$ & $3$ samples \\\hline
         Uplink pilot period, $\tpl$ & \revise{$10$ or $20$} samples \\\hline
         Downlink data period, $\td$ & $\!\frac12(\tc \!-\! \tpl \!-\! 2\tg)$ samples \\\hline
         AP/UE antenna height & $10/1.5\m$ \\ \hline
         UE's transmit power, $\bar \rho\sub{UE}$ & $100 \mW$ \\\hline
         AP's transmit power, $\bar \rho\sub{AP}$ & $200 \mW$ \\\hline
         Noise power & $-94\dBm$ \\ \hline
         Phase noise spectrum level, $\revise{S_{100}}\!\!$ & between $[-120,-80]\dBcHz\!$ \\ \hline
    \end{tabular}
\end{table}
The LO quality constant $c_\nu$ is computed from the phase noise spectrum level $\revise{S_{100}}$ as in~\eqref{eq:PN_spectrum_level}. 

 We let the APs send a downlink pilot at the first sample of the downlink period, i.e., at time $\tpl + \tu + \tg + 1$, of every slot.  
 The expectation $\Exop[\Delta_{k,\ell,i}]$ is computed over $10^3$ realizations of consecutive frames, and the \gls{CDF} or mean of the \gls{SE} is obtained over $200$ realizations of the AP/UE locations and shadow fading profiles.


\subsection{\revise{The Two-AP Case}} \label{sec:result_2AP}

\revise{We first assume that two APs are present in the simulation area and they are placed $250\m$ apart. Note that, because the area is wrapped around, the exact positions of the two APs are immaterial.}

\begin{figure}[t!]
    \centering
    \includegraphics[clip,trim={.5cm 0cm 1cm .4cm},width=\linewidth]{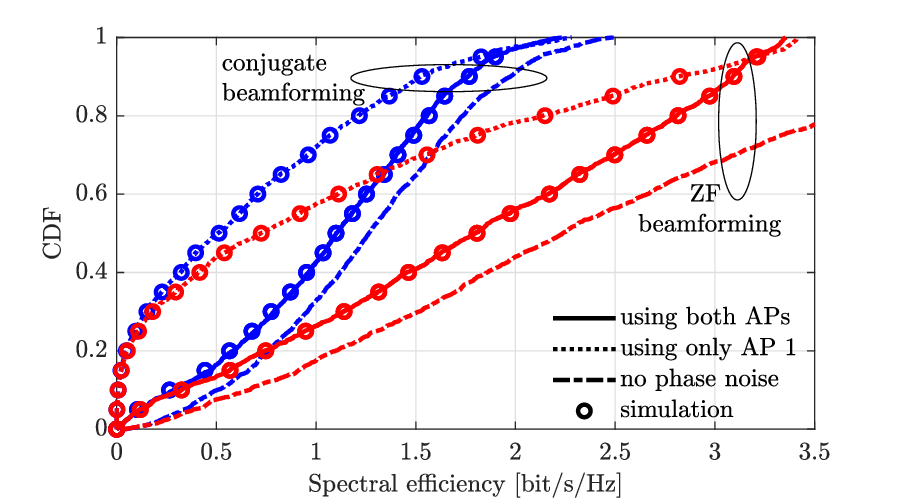}
    \caption{\Gls{CDF} of the per-UE downlink SE for $L = 2$~APs, $N = 64$ antennas per AP, $K = 10$ UEs,  \revise{$\tpl = 10$ samples,} frame length $\nsync = 1$ slots, and phase noise spectrum level $\revise{S_{100}} = -80\dBcHz$. \revise{Lines depict the spectral efficiency computed using the closed-form formulas in~\eqref{eq:rate_i_conj}~\eqref{eq:rate_i_ZF}; circle markers depict the \gls{SE} computed using~\eqref{eq:rate_i} with Monte-Carlo simulation of $|\DS_{k,i}|$, $\Exop[|\BU_{k,i}|^2]$, and $\Exop[|\UI_{k,i}|^2]$}.
    }
    \label{fig:two-AP}
\end{figure}

\revise{In Fig.~\ref{fig:two-AP}, we present the \gls{CDF} of the per-UE downlink SE computed using Corollary~\ref{cor:SE} for 
frame length $\nsync = 1$ slots, and phase noise spectrum level $\revise{S_{100}} = -80\dBcHz$.} 
Phase estimation is performed using the designed Kalman filter. 
    We also show 
    two benchmarks: the case with no phase noise and the case with AP~$2$ turned off. In the latter, AP~$1$ still transmits a demodulation pilot in every slot, but no inter-AP phase calibration is needed. We observe that
    phase noise substantially degrades the \gls{SE} compared to the no-phase-noise benchmark. 
    Furthermore, when there is phase noise, coherent beamforming from both APs with calibration
    significantly outperforms the single-AP scenario, especially for UEs in unfavorable locations. Compared to conjugate beamforming, \gls{ZF} beamforming performs substantially better, but its SE loss due to phase noise is more significant.\footnote{\revise{In some specific  settings (e.g., with a limited number of AP antennas), not reported in this paper, conjugate beamforming can achieve higher SE than \gls{ZF} beamforming. {However, ZF beamforming is generally preferable.}}} \revise{Finally, we also plot the \gls{SE} computed using~\eqref{eq:rate_i} with Monte-Carlo simulation of $|\DS_{k,i}|$, $\Exop[|\BU_{k,i}|^2]$, and $\Exop[|\UI_{k,i}|^2]$ with the formulas in~\eqref{eq:tmp410}--\eqref{eq:tmp418}. The simulation results match closely the values obtained with the closed-form expressions in~\eqref{eq:rate_i_conj}, \eqref{eq:rate_i_ZF}, validating these expressions.}
%

\begin{figure}[t!]
    \centering
    \includegraphics[clip,trim={.4cm 0cm 1cm .4cm},width=\linewidth]{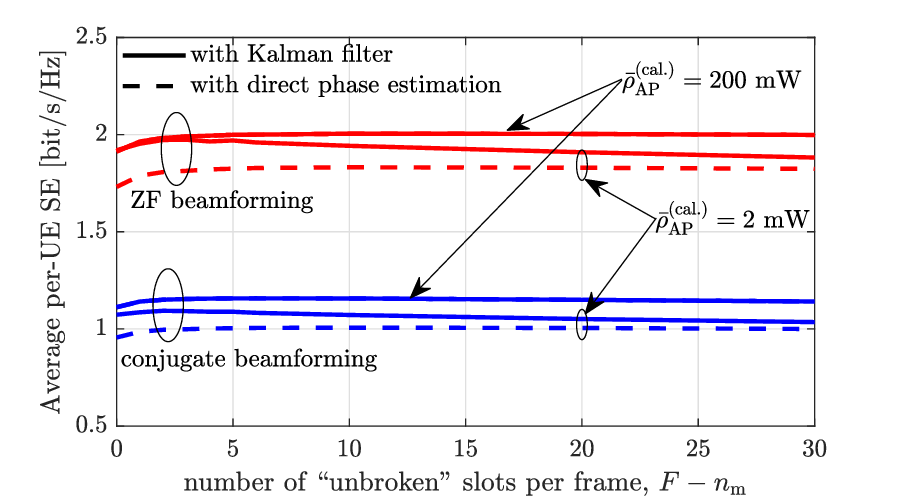}
    \caption{\revise{Average per-UE downlink SE (achieved with the Kalman filter) vs. the number of ``unbroken'' slots per frame, $F - 1$, for $L =2$~APs, $N = 64$ antennas per AP, $K = 10$ UEs, $\tpl = 10$ samples, phase noise spectrum level $\revise{S_{100} = -90 \dBcHz}$, and calibration signal transmit power $\bar{\rho}\of{\rm cal.}\sub{AP} \mW$}.}
    \label{fig:avgSE_vs_framelength_-90dBcHz}
\end{figure}
\begin{figure}[t!]
    \centering
    \includegraphics[clip,trim={.4cm 0cm 1cm .4cm},width=\linewidth]{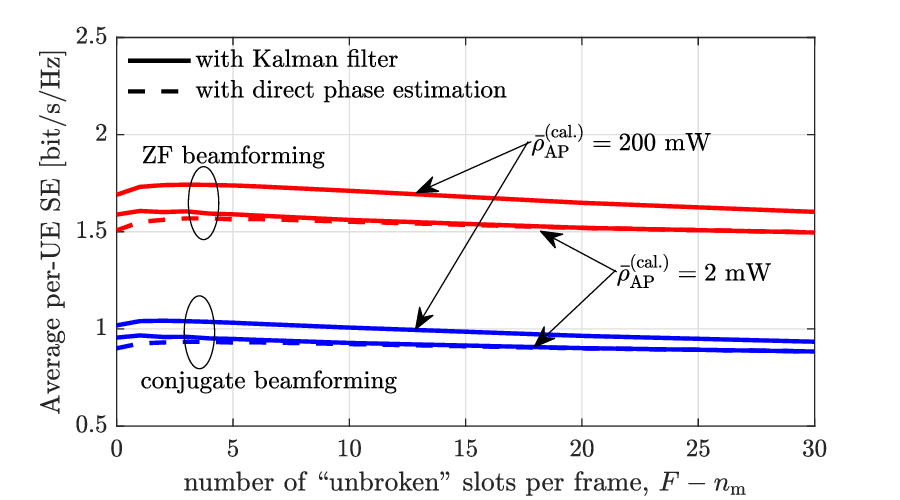}
    \caption{\revise{Similar to Fig.~\ref{fig:avgSE_vs_framelength_-90dBcHz} but with $S_{100} = -80 \dBcHz$.}}
    \label{fig:avgSE_vs_framelength_-80dBcHz}
\end{figure}

Next, we investigate the impact of the frequency of phase measurements. We start with the minimal frame length $F = 1$, i.e., phase measurement is done in every slot, and progressively insert more ``unbroken'' slots in which there is no phase measurement, \revise{and thus phase calibration is done based on measurements in previous slots}. The unbroken slots are placed evenly within a frame and across frames. \revise{In Fig.~\ref{fig:avgSE_vs_framelength_-90dBcHz}, we show the average per-UE SE as a function of the number of unbroken slots, $F - 1$, for $S_{100} = -90\dBcHz$ and other parameters set similarly as in Fig.~\ref{fig:two-AP}. Phase calibration is done with the direct estimation $\bar\alpha_{i_2}$ in~\eqref{eq:2AP_phi_hat} or the Kalman filter output~\eqref{eq:2AP_update}. We assume that the calibration signals are transmitted with power $\bar{\rho}\of{\rm cal.}\sub{AP} = 200 \mW$ (equal to the downlink data transmit power) or $\bar{\rho}\of{\rm cal.}\sub{AP} = 2 \mW$ ($1\%$ of the downlink data transmit power). We observe that the average SE initially increases and then gradually decreases as the frame length grows. This behavior reflects a tradeoff: shorter frames enable more frequent phase calibration, while longer frames reduce service interruptions of AP~$2$. Furthermore, the improvement brought by the Kalman filter is visible when the calibration signals are transmitted with low power, leading to low-quality direct estimation. This agrees with our remark at the end of Section~\ref{sec:kalman}.

In Fig.~\ref{fig:avgSE_vs_framelength_-80dBcHz}, we plot the average per-\gls{UE} \gls{SE} for $S_{100} = -80\dBcHz$, i.e., the LOs drift more rapidly. Similar observations as in Fig.~\ref{fig:avgSE_vs_framelength_-90dBcHz} hold, except that the \gls{SE} degrades more quickly as $\nsync$ grows, and the gain from using the Kalman filter is less significant.
}



\subsection{\revise{The General Case}}
\revise{We now consider $L \ge 2$ \glspl{AP} and assume that the APs are located uniformly at random within the area, such that they are at least $10\m$ apart. We also keep $F = \nmeas$ slots. We allocate powers for the calibration signals transmitted from the master AP as in~\eqref{eq:FPA_sync}. 
Unless stated otherwise, we let the graph $\setG$ contain $M = L-1$ edges such that the sum of the channel strength $\|\matG_{ij}\|\sub{F}$ over the edges $(i,j)$ is maximized. That is, $\setG$ is constructed as the minimum spanning tree of a graph connecting all nodes and have edge weights $-\|\matG_{i,j}\|\sub{F}$.   
We color $\setG$ using an adaptation of the greedy Welsh–Powell algorithm~\cite{Welsh67}.}

\begin{figure}[t!]
    \centering
    \includegraphics[clip,trim={.4cm 0cm 1cm .4cm},width=\linewidth]{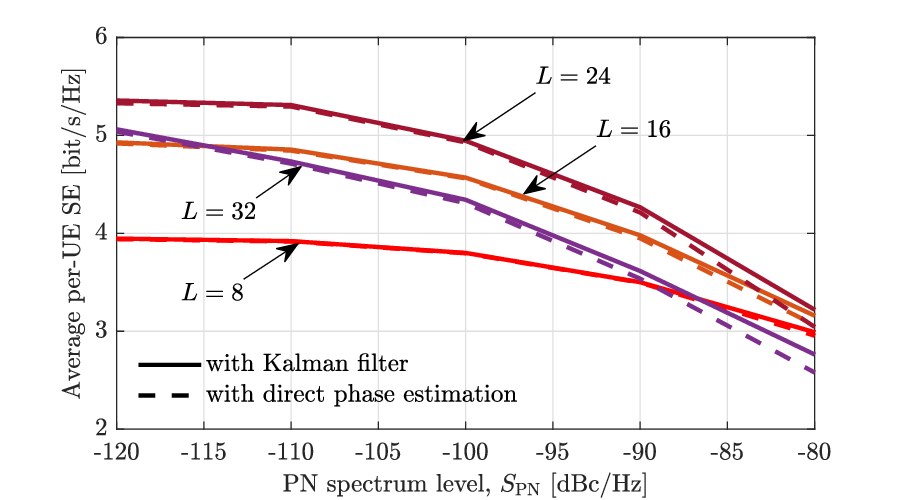}
    \caption{Average per-UE downlink SE vs. the PN spectrum level $\revise{S_{100}}$ for ZF beamforming with \revise{$L \in \{8,16,24,32\}$}~APs, $N = 64$ antennas per AP, $K = 10$ UEs, \revise{ $\tpl = 10$ samples,} frame length $\nsync = \nmeas$ slots, and $M = L - 1$ edges.}
    \label{fig:avgSE_vs_PN}
\end{figure}

Fig.~\ref{fig:avgSE_vs_PN} depicts the average per-UE downlink SE as a function of the LOs' quality for a typical range of the PN spectrum level $\revise{S_{100}}$ from $-120$ to $-80 \dBcHz$. We consider $\revise{L \in \{8,16,24,32\}}$~APs and ZF beamforming (the results for conjugate beamforming follow the same trend). 
We show the SE achieved with both the Kalman filter and the direct phase estimation~\eqref{eq:measure_l1_l2}. The gain brought by the Kalman filter is more pronounced for a higher number of APs and lower LO quality in this range. Furthermore, as expected, the SE deteriorates as the LOs' quality worsens. Remarkably, the advantage of using more APs diminishes for low-quality LOs. \revise{For example,  using $32$ APs leads to a lower SE than using $16$ APs for this range of phase noise spectrum.}  

\begin{figure}[t!]
    \centering
    \includegraphics[clip,trim={.5cm 0cm 1cm .4cm},width=\linewidth]{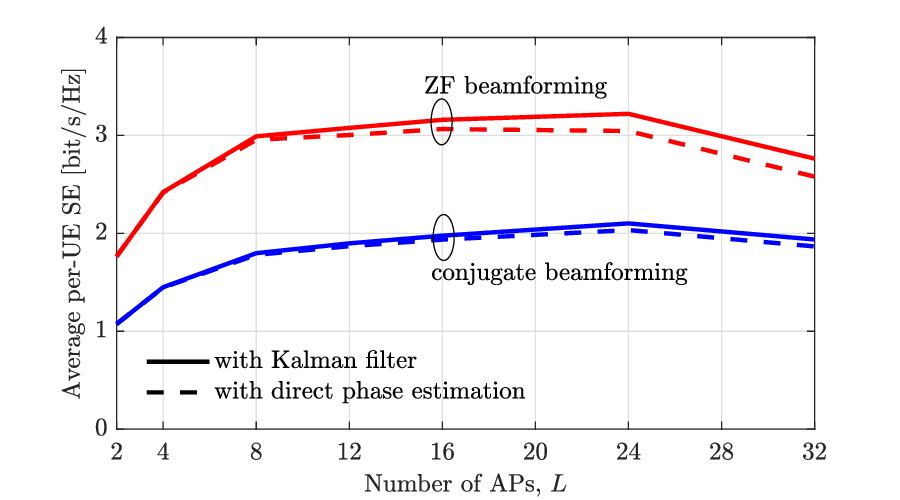}
    \caption{Average per-UE downlink SE for $L\in \revise{[2:32]}$ APs, $N = 64$ antennas per AP, $K = 10$ UEs, \revise{ $\tpl = 10$ samples,} PN spectrum level $\revise{S_{100}} = -80 \dBcHz$, frame length $\nsync = \nmeas$ slots, and $M = L - 1$ edges.}
    \label{fig:various_L}
\end{figure}

To further examine the impact of using more APs, in Fig.~\ref{fig:various_L}, we fix $\revise{S_{100}} = -80\dBcHz$ and plot the average per-UE downlink SE as a function of $L$ while keeping other parameters similar to Fig.~\ref{fig:avgSE_vs_PN}. 
As $L$ increases, the SE first improves for both conjugate and ZF beamforming. However, beyond a certain number of APs \revise{(namely, 24 \glspl{AP} for this setting)}, the achievable SE begins to degrade. This is because, when there are more APs, $\nmeas$ increases. The measurements of different components of $\vecalpha$ are spread out over more slots, raising the offset noise $\vecxi_n$ (see~\eqref{eq:kalman_observation_manyAP}). 
This result highlights a tradeoff between spatial diversity and calibration overhead when deploying more~APs. 
%
%

A natural question arises: 
should the graph $\setG$ contain more links to allow for more measurements? 
In our experiments, however, increasing $M$ from its minimum required value 
$L-1$ 
leads to lower SE. 
This is because, while keeping more edges in $\setG$ leads to more phase measurements, this entails 1) a possibly larger number of colors and thus a larger number of necessary measurement slots, $n\sub{m}$, and 2) a larger number of \revise{calibration} signals that share the power budget $\Pdl$ in~\eqref{eq:sync_signal_masterAP}. This emphasizes that when the cost of phase measurements is taken into account, it can be beneficial to limit the number of measurements. 
%

\begin{figure}[t!]
    \centering
    \includegraphics[clip,trim={.5cm 0cm 1cm .4cm},width=\linewidth]{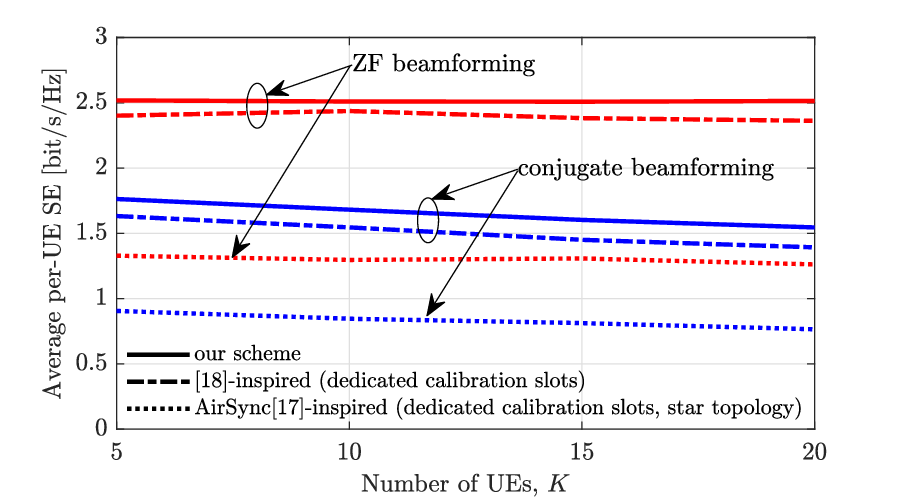}
    \caption{\revise{Average per-UE downlink SE of the proposed scheme vs. the number of \glspl{UE}, $K$, compared to schemes inspired by~\cite{Balan13,Rogalin14} with dedicated calibration slots, for $L = 24$ APs, $N = 64$ antennas per AP, $\tpl = 20$ samples, $K \in [5:20]$ UEs, PN spectrum level $S_{100} = -80 \dBcHz$, frame length $\nsync = \nmeas$ slots, and $M = L - 1$ edges.}}
    \label{fig:various_K}
\end{figure}

\revise{
Finally, in Fig.~\ref{fig:various_K}, we plot the achievable average per-UE downlink SE as a function of the number of UEs, $K$, for $L = 24$ \glspl{AP}. We set $\tpl = 20$ and $K \in [5:20]$, so that there is no pilot contamination.\footnote{\revise{Recall that we assume no pilot contamination, to focus on the impact of phase noise.}} The average SE achieved with \gls{ZF} beamforming remains stable as $K$ grows, while that with conjugate beamforming decreases gradually. We also compare with baseline schemes inspired by~\cite{Balan13,Rogalin14} where calibration signal transmissions are scheduled between the APs in a graph-based manner similar to our scheme, but placed in dedicated slots; see Fig.~\ref{fig:TDD_operation}(b). For these baselines, we consider two topologies for the graph $\setG$: i) a star topology, inspired by AirSync~\cite{Balan13}, where the the AP with the largest sum \gls{SNR} to other APs is choosen as the master AP; ii) a minimum-spanning-tree based construction of $\setG$ (as in our scheme), inspired by~\cite{Rogalin14}. Furthermore, as in~\cite{Balan13,Rogalin14}, we do not account for LO drifts in these baselines. Our scheme outperforms both baselines, especially the AirSync-inspired scheme with the star topology, where no two APs can transmit calibration signals simultaneously. 
}


    






\section{Conclusions} \label{sec:conclusion}

We showed the feasibility of disrupting the conventional \gls{TDD} flow in distributed antenna systems to enable over-the-air bidirectional phase measurements for phase calibration among \glspl{AP} with independent \glspl{LO}. We developed a Kalman filter for tracking inter-AP phase shifts based on these measurements, and analyzed the achievable \gls{SE} of downlink beamforming when using the estimated phase shifts for coherent transmission. \revise{Our scheme, with phase calibration signals interleaved into the TDD flow, leads to higher achievable downlink SE than the scheme with dedicated calibration slots.} Our results demonstrated that frequent phase reestimation is important in systems with rapidly drifting \glspl{LO}, and 
that increasing the number of \glspl{AP} and the amount of phase measurements between them can degrade the \gls{SE} due to the growing overhead induced by the measurements. 


\begin{appendices}
\revise{
\section{Generalized Kalman Filter} \label{app:Kalman}
    The conventional Kalman filter~\cite[Chap.~5]{Simon2006optimal} assumes that the process noise and measurement noise are uncorrelated. A generalized Kalman filter for correlated process and measurement noise was derived in~\cite[Sec.~7.1]{Simon2006optimal}. For completeness, below we describe  this generalized filter.

    Consider process and measurement models
    \begin{align}
        \vecx_n &= \matF_{n-1} \vecx_{n-1} + \vecw_{n-1}, \\
        \vecy_n &= \matA \vecx_n + \vecv_n
    \end{align}
    where $\vecw_n$ and $\vecv_n$ are independent across~$n$ and distributed as $\normal(\veczero, \matQ_n)$ and $\normal(\veczero, \matR_n)$, respectively. Assume that $\vecw_n$ and $\vecv_n$ are correlated as
    \begin{equation}
        \Exop[\vecw_n \tp{\vecv}_{j}] = \begin{cases}
            \matM_n, &\text{if~} j = n + 1, \\
            \matzero, &\text{otherwise}.
        \end{cases}
    \end{equation}
    The generalized Kalman filter is initialized with the posterior estimate $\vecx_0^+$ and posterior error covariance matrix $\matP_0^+$. Then, for each time step $n = 1,2,\dots$, the generalized Kalman filter equations are given as follows.
    \begin{itemize}
        \item Prior error covariance matrix update: 
        \begin{equation}
            \matP_n^- = \matF_{n-1} \matP_{n-1}^+ \tp{\matF}_{n-1} + \matQ_{n-1}; \label{eq:Kalman_prior-update}
        \end{equation}

        \item Kalman gain update:
        \begin{align}
            \matK_n = (\matP_n^+ \tp{\matA}_n)(\matA_n \matP_n^- \tp{\matA}_n + {\matA}_n {\matM}_n + \tp{\matM}_n \tp{\matA}_n + \matR_n)^{-1};
        \end{align}

        \item Prior estimate update: 
        \begin{equation}
            \vecx^-_n = \matF_{n-1} \vecx^+_{n-1};
        \end{equation}

        \item Posterior estimate update:
        \begin{equation}
            \vecx^+_n = \vecx^-_n + \matK_n(\vecy_n - \matA_n \vecx^-_n);
        \end{equation}

        \item Posterior error covariance matrix update:
        \begin{equation}
            \matP^+_n = \matP_n^- - \matK_n (\matA_n \matP^-_n + \tp{\matM}_n). \label{eq:Kalman_posterior_update}
        \end{equation}
    \end{itemize}
}

\section{Computation of the Covariance/Correlation Matrices for the Kalman Filter} \label{app:cov}

Within a frame, we update the estimate of $\vecalpha$ at time $i_2$ of each measurement slot. In the following, we focus on the $n$th measurement slot of a generic frame, and thus drop the index~$n$ for notational simplicity. Let $i$ be the global index of time~$i_2$ of this measurement slot. Let $d$ be such that $d - 1$ is the number of ``unbroken'' slots between this measurement slot and the previous one. We wrap the frame around so that the $(-1)$th measurement slot is the $\nmeas$th measurement slot of the previous frame.  Recall that each element of $\vecalpha$ is identified with an AP pair $(\ell_1,\ell_2)$.

\subsection{Computation of $\matSigma\of{\veczeta}$}
\label{sec:compute_cov_zeta}
The element of $\veczeta$ corresponding to the AP pair $(\ell_1,\ell_2)$ is 
\revise{
\begin{align}
    &[\nut_{\ell_2,i} + \nur_{\ell_2,[i]_{\floor{K/2}}} - \nut_{\ell_1,i} - \nur_{\ell_1,[i]_{\floor{K/2}}}] \notag \\
    &\quad- [\nut_{\ell_2,i-d \tc} \!+\! \nur_{\ell_2,[i]_{\floor{K/2}}-d \tc} \!-\! \nut_{\ell_1,i-d \tc} \!-\! \nur_{\ell_1,[i]_{\floor{K/2}}-d \tc}] \notag \\
    &= [\nut_{\ell_2,i} - \nut_{\ell_2,i - d \tc}] + [\nur_{\ell_2,[i]_{\floor{K/2}}} - \nur_{\ell_2,[i]_{\floor{K/2}}-d \tc}] \notag \\
    &\quad - [\nut_{\ell_1,i} \!-\! \nut_{\ell_1,i - d \tc}] - [\nur_{\ell_1,[i]_{\floor{K/2}}} \!-\! \nur_{\ell_1,[i]_{\floor{K/2}}-d \tc}],
\end{align}}
which is the drift of \revise{$\nut_{\ell_1}$ and $\nut_{\ell_2}$} from $i - d \tc$ to $i$ 
and of \revise{$\nur_{\ell_1}$ and $\nur_{\ell_2}$} from $[i]_{\floor{K/2}} - d \tc$ to $[i]_{\floor{K/2}}$. 
By summing the PN variance over these intervals, we obtain the diagonal element of $\matSigma\of{\veczeta}$ corresponding to  $(\ell_1,\ell_2)$ as
\revise{$4d\tau_c \sigma_\nu^2$}. The off-diagonal element of $\matSigma\of{\veczeta}$ corresponding to AP pairs $(\ell_{1a},\ell_{2a})$ and $(\ell_{1b},\ell_{2b})$ is nonzero only if the pairs share a common AP, and given by \revise{$\varsigma \cdot 2d\tau_c\sigma_\nu^2$} with 
\begin{equation} \label{eq:varsigma}
    \varsigma = \begin{cases}
        1, &\text{if $\ell_{1a} = \ell_{1b}$ or $\ell_{2a} = \ell_{2b}$}, \\
        -1, &\text{if $\ell_{1a} = \ell_{2b}$ or $\ell_{2a} = \ell_{1b}$}.
    \end{cases}
\end{equation}



\subsection{Computation of $\matSigma\of{\vecxi}$}
\label{sec:compute_cov_xi}
Consider a connected AP pair $(\ell_1,\ell_2)$ where one of them is a master AP in the current slot.
Let $i_{\ell_1}$ and $i_{\ell_2}$ be the indices of the latest sample times, up to $i$, when calibration signal is sent from AP~$\ell_2$ to AP~$\ell_1$ and from AP~$\ell_1$ to AP~$\ell_2$, respectively. 
 \revise{The time $\max\{i_{\ell_1}, i_{\ell_2}\}$ coincides with either time $i_1$ or $i_2$ of the current slot. (In the two-AP case, $(i_{\ell_1}, i_{\ell_2}) = (i_1, i_2)$.)} 
The element of $\vecxi$ corresponding to $(\ell_1, \ell_2)$ is given by
\revise{
\begin{align}
    &[\nut_{\ell_2,i_{\ell_1}} + \nur_{\ell_2,i_{\ell_2}} - \nur_{\ell_1,i_{\ell_1}} - \nut_{\ell_1,i_{\ell_2}}] \notag \\
    &\qquad- [\nut_{\ell_2,i} + \nur_{\ell_2,[i]_{\floor{K/2}}} - \nut_{\ell_1,i} - \nur_{\ell_1,[i]_{\floor{K/2}}}] \notag \\
    &= [\nut_{\ell_1,i} - \nut_{\ell_1,i_{\ell_2}}]  + [\nur_{\ell_1,[i]_{\floor{K/2}}} - \nur_{\ell_1,i_{\ell_1}}] \notag \\
    &\qquad - 
    [\nut_{\ell_2,i} - \nut_{\ell_2,i_{\ell_1}}] -
    [\nur_{\ell_2,[i]_{\floor{K/2}}} - \nur_{\ell_2,i_{\ell_2}}]. \label{eq:tmp1788}
\end{align}
which contains the drift of $\nut_{\ell_1}$ from $i_{\ell_2}$ to $i$, of $\nur_{\ell_1}$ from $i_{\ell_1}$ to $[i]_{\lfloor K/2 \rfloor}$, of $\nut_{\ell_2}$ from $i_{\ell_1}$ to $i$,  and of $\nur_{\ell_2}$ from $i_{\ell_2}$ to $[i]_{\lfloor K/2 \rfloor}$.} 
By summing the phase noise variance across these intervals, we obtain the diagonal element of $\matSigma\of\vecxi$ corresponding to $(\ell_1, \ell_2)$ as
\revise{
    \begin{align}
        &(|i-i_{\ell_2}| + |[i]_{\lfloor K/2 \rfloor} - i_{\ell_1}| + |i-i_{\ell_1}| + |[i]_{\lfloor K/2 \rfloor} - i_{\ell_2}|)\sigma_\nu^2 \notag \\
        &= 2(i - \min\{i_{\ell_1}, i_{\ell_2}, [i]_{\lfloor K/2 \rfloor}\}) \sigma_\nu^2. 
    \end{align}
    }
    
    The off-diagonal element corresponding to $(\ell_{1a}, \ell_{2a})$ and $(\ell_{1b}, \ell_{2b})$ is nonzero only if the two pairs share a common AP. \revise{By inspecting the variance of the repeating terms in~\eqref{eq:tmp1788} for $(\ell_1,\ell_2) = (\ell_{1a},\ell_{2a})$ and $(\ell_1,\ell_2) = (\ell_{1b},\ell_{2b})$, we express the mentioned off-diagonal element of $\matSigma\of{\vecxi}$ as
    \begin{align}
        \begin{cases}
            &[i - \max\{i_{\ell_{2a}}, i_{\ell_{2b}}\} + \iota([i]_{\floor{K/2}}, i_{\ell_{1a}}, i_{\ell_{1b}})] \sigma_\nu^2, \\
            &\qquad\qquad\text{if~} \ell_{1a} = \ell_{1b}, \\ 
            &[i - \max\{i_{\ell_{1a}}, i_{\ell_{1b}}\} + \iota([i]_{\floor{K/2}}, i_{\ell_{2a}}, i_{\ell_{2b}})] \sigma_\nu^2, \\
            &\qquad\qquad\text{if~} \ell_{2a} = \ell_{2b}, \\ 
            &-[i - \max\{i_{\ell_{1b}}, i_{\ell_{2a}}\} + \iota([i]_{\floor{K/2}}, i_{\ell_{1a}}, i_{\ell_{2b}})] \sigma_\nu^2, \\
            &\qquad\qquad\text{if~} \ell_{1a} = \ell_{2b}, \\ 
            &-[i - \max\{i_{\ell_{1a}}, i_{\ell_{2b}}\} + \iota([i]_{\floor{K/2}}, i_{\ell_{1b}}, i_{\ell_{1a}})] \sigma_\nu^2, \\
            &\qquad\qquad\text{if~} \ell_{1b} = \ell_{2a}, 
        \end{cases}    \end{align}
        with 
        \begin{align}
            \iota(a,b,c) = \ind{a - \max\{b,c\}} + \ind{\min\{b,c\} - a}. 
        \end{align}
        }

\subsection{Computation of $\matSigma\of{\veczeta\vecxi}$}
\label{sec:compute_corr}
Recall that $\matSigma\of{\veczeta\vecxi} = \Exop[\veczeta \tp{\vecxi}]$.
\revise{Notice that the element of $\veczeta$ corresponding to $(\ell_1,\ell_2)$ subsumes the negation of the element of $\vecxi$ corresponding to $(\ell_1,\ell_2)$. Therefore, 
}
the diagonal element of $\matSigma\of{\veczeta\vecxi}$ corresponding to $(\ell_1,\ell_2)$ is given by
\revise{
\begin{equation}
    \sigma^2_{\veczeta\vecxi}(\ell_1,\ell_1) + \sigma^2_{\veczeta\vecxi}(\ell_2,\ell_2)
\end{equation}
where
\begin{align}
    \sigma^2_{\veczeta\vecxi}(x,y) &= - [i - \max\{i_{x}, i - d\tc\}  \notag \\
    &\quad +  \iota([i]_{\floor{K/2}}, i_{y}, [i]_{\floor{K/2}} - d \tc)] \sigma_\nu^2.
\end{align}
}
The off-diagonal element corresponding to $(\ell_{1a}, \ell_{2a})$ and $(\ell_{1b}, \ell_{2b})$ is nonzero only if the two pairs share a common AP, and given by 
\revise{
\begin{equation}
    \begin{cases}
        \varsigma \sigma^2_{\veczeta\vecxi}(\ell_{1b},\ell_{2b}), &\text{if $\ell_{1a} = \ell_{2b}$ or $\ell_{2a} = \ell_{2b}$}, \\
        \varsigma \sigma^2_{\veczeta\vecxi}(\ell_{2b},\ell_{1b}), &\text{if $\ell_{1a} = \ell_{1b}$ or $\ell_{2a} = \ell_{1b}$}
    \end{cases}
\end{equation}
}
with $\varsigma$ defined in~\eqref{eq:varsigma}. 
    
\section{Proof of Theorem~\ref{th:rate_i}} \label{proof:rate_i}

In the following, we omit the subscript $i$ in $\vecq_{k,\ell,i}$ and $\hat \vecq_{k,\ell,i}$ for notational simplicity.
We will use the following lemma. 

\begin{lem} \label{lem:h_estimate_property}
    It holds that 
    \begin{enumerate}
        \item $\hat \vecq_{k,\ell} \sim \jpg\left(\veczero, \gamma_{k,\ell}\matidentity_N\right)$  with $\gamma_{k,\ell} = \sqrt{\Ppl K} \beta_{k,\ell} c_{k,\ell}$;
        
        \item $\Exop[\tp{\tilde \vecq}_{k,\ell} \hat \vecq_{k,\ell}^* \given \nu_{\ell,[i]_k}] = 0$ where $\tilde \vecq_{k,\ell} =  \vecq_{k,\ell} - \hat \vecq_{k,\ell} \sim  \jpg\left(\veczero, \beta_{k,\ell} - \gamma_{k,\ell}\matidentity_N\right)$ is the channel estimation error;

        
        \item $\Exop[|\tp{\tilde \vecq}_{k,\ell} \hat\vecq_{k,\ell}^*|^2] = N \gamma_{k,\ell}(\beta_{k,\ell} - \gamma_{k,\ell})$;
        
        \item $\Exop[\|\hat\vecq_{k,\ell}\|^4] = N(N + 1)\gamma_{k,\ell}^2$.
    \end{enumerate}
    
\end{lem}
\begin{proof}
    Conditioned on $\nu_{\ell,[i]_k}$, $\hat{\vecq}_{k,\ell}$ is the \gls{MMSE} estimate of $\vecq_{k,\ell}$. The lemma follows directly from properties of the \gls{MMSE} estimate and some simple manipulations.
\end{proof}



The derivation of the achievable rate follows similar steps as in~\cite[Chap.~3]{marzetta2016fundamentals} and~\cite{Ngo17_CF}, but with the extra factor $\Delta_{k,\ell,i}$.

\subsection{Conjugate Beamforming}

\subsubsection{Compute $\DS_{k,i}$} We proceed as
\begin{align}
    &\DS_{k,i} \notag \\
    &= \sqrt{\Pdl} \sum_{\ell = 1}^L a_{\ell,i} \sqrt{\frac{\eta_{k,\ell}}{N \gamma_{k,\ell}}} \Exop\left[\Delta_{k,\ell,i} \tp{\vecq_{k,\ell}} \hat\vecq_{k,\ell}^* \right] \\
    &= \sqrt{\Pdl} \sum_{\ell = 1}^L a_{\ell,i} \sqrt{\frac{\eta_{k,\ell}}{N \gamma_{k,\ell}}} \Exop\!\Big[\Delta_{k,\ell,i} \Exop\!\big[\tp{\vecq_{k,\ell}} \hat\vecq_{k,\ell}^* \given \nu_{\ell,[i]_k}\big] \Big]  \label{eq:tmp258}\\
    &= \sqrt{\Pdl} \sum_{\ell = 1}^L a_{\ell,i} \sqrt{\eta_{k,\ell} N \gamma_{k,\ell}} \Exop\left[\Delta_{k,\ell,i}\right], \label{eq:DS}
\end{align} 
where \eqref{eq:tmp258} holds because given $\nu_{\ell,[i]_k}$, $\tp{\vecq_{k,\ell}} \hat\vecq_{k,\ell}^*$ is independent of $\Delta_{k,\ell,i}$, and \eqref{eq:DS} follows from $\Exop\left[\tp{\vecq_{k,\ell}} \hat\vecq_{k,\ell}^* \given \nu_{\ell,[i]_k} \right] = \Exop\left[\|\hat\vecq_{k,\ell}\|^2 \given \nu_{\ell,[i]_k} \right] = N \gamma_{k,\ell}$.

\subsubsection{Compute $\Exop\left[\abs{\BU_{k,i}}^2\right]$} 
Notice that the variables $\Delta_{k,\ell,i} \tp{\vecq}_{k,\ell} \hat\vecq_{k,\ell}^*$ are uncorrelated across $\ell$.
Using the fact that the variance of the sum of uncorrelated random variables is equal to the sum of the variances, we have that 
\begin{align}
    &\Exop\left[\abs{\BU_{k,i}}^2\right] \notag \\
    &= {\Pdl}\sum_{\ell = 1}^L a_{\ell,i} \frac{\eta_{k,\ell}}{N \gamma_{k,\ell}} \Exop\!\left[\left|\Delta_{k,\ell,i} \tp{\vecq}_{k,\ell} \hat\vecq_{k,\ell}^* - \Exop\!\left[\Delta_{k,\ell,i} \tp{\vecq}_{k,\ell} \hat\vecq_{k,\ell}^*\right] \right|^2\right] \\ 
    &= {\Pdl}\sum_{\ell = 1}^L a_{\ell,i} \frac{\eta_{k,\ell}}{N \gamma_{k,\ell}} \Big(\underbrace{\Exop\left[\left|\Delta_{k,\ell,i} \tp{\vecq}_{k,\ell} \hat\vecq_{k,\ell}^*\right|^2\right]}_{A} \notag \\
    &\qquad - \underbrace{\left|\Exop\left[\Delta_{k,\ell,i} \tp{\vecq}_{k,\ell} \hat\vecq_{k,\ell}^*\right] \right|^2}_{B} \Big). \label{eq:tmpBU}
\end{align}
The same computation leading to~\eqref{eq:DS} shows that $B = \left|\Exop\left[\Delta_{k,\ell,i}\right]\right|^2 N^2 \gamma_{k,\ell}^2$. We compute the term $A$ as follows
\begin{align}
    A &= \Exop_{\nu_{\ell,[i]_k}}\left[\Exop\left[\left|\tp{\tilde \vecq}_{k,\ell} \hat\vecq_{k,\ell}^* +  \|\hat\vecq_{k,\ell}\|^2\right|^2 \given \nu_{\ell,[i]_k}\right]\right] \\ 
    &= \Exop\left[\left|\tp{\tilde \vecq}_{k,\ell} \hat\vecq_{k,\ell}^*\right|^2\right] +  \Exop\left[\|\hat\vecq_{k,\ell}\|^4\right] \label{eq:tmp280} \\ 
    &= N \gamma_{k,\ell} (\beta_{k,\ell} - \gamma_{k,\ell}) + N(N + 1) \gamma_{k,\ell}^2 \label{eq:tmp293} \\ 
    &= N \gamma_{k,\ell} \beta_{k,\ell} + N^2 \gamma_{k,\ell}^2,
\end{align}
where~\eqref{eq:tmp280} follows since given $\nu_{\ell,[i]_k}$, $\tilde \vecq_{k,\ell}$ has mean $\veczero$ and is uncorelated with $\hat\vecq_{k,\ell}$, and~\eqref{eq:tmp293} follows from Lemma~\ref{lem:h_estimate_property}.
Using the computation of $A$ and $B$ in~\eqref{eq:tmpBU}, we obtain
\begin{multline}
    \Exop\left[\abs{\BU_{k,i}}^2\right] = \\ {\Pdl}\sum_{\ell = 1}^L a_{\ell,i} \eta_{k,\ell} \big( \beta_{k,\ell} 
    + N \gamma_{k,\ell} \big(1 - \left|\Exop\left[\Delta_{k,\ell,i}\right]\right|^2\big) \big). \label{eq:BU}
\end{multline} 

\subsubsection{Compute $\Exop\left[\abs{\UI_{k,i}}^2\right]$} As the variables $\UI_{k,k',i}$ are uncoorelated across $k'$, we have that $ \Exop\left[\abs{\UI_{k,i}}^2\right] = \sum_{k' = 1, k'\ne k}^K \Exop\left[\abs{\UI_{k,k',i}}^2\right]$. Next, by expanding $\hat\vecq_{k',\ell}^*$, we obtain
\begin{align}
    &\Exop\left[\abs{\UI_{k,k',i}}^2\right]
    \notag \\
    &= \Exop\Bigg[\bigg|\sqrt{\Pdl} \sum_{\ell = 1}^L a_{\ell,i}  \sqrt{\frac{\eta_{k',\ell}}{N \gamma_{k',\ell}}} \Delta_{k,\ell,i} \notag \\
    & \qquad \cdot \tp{\vecq}_{k,\ell} c_{k',\ell}\Big(\sqrt{\Ppl K} \vecq_{k',\ell} + \vecz\supp{pilot}_{k',\ell}\Big)\bigg|^2\Bigg] \\ 
    &= \Pdl \Exop\Bigg[\bigg|\sum_{\ell = 1}^L a_{\ell,i} \sqrt{\frac{\eta_{k',\ell}}{N \gamma_{k',\ell}}} \Delta_{k,\ell,i} c_{k',\ell}\sqrt{\Ppl K}  
    \tp{\vecq}_{k,\ell}  \vecq_{k',\ell}\bigg|^2\Bigg] \notag \\
    &\quad + \Pdl \Exop\!\Bigg[\bigg|\sum_{\ell = 1}^L a_{\ell,i}  \sqrt{\frac{\eta_{k',\ell}}{N \gamma_{k',\ell}}} \Delta_{k,\ell,i} c_{k',\ell} 
    \tp{\vecq}_{k,\ell} \vecz\supp{pilot}_{k',\ell}\bigg|^2\Bigg] \label{eq:tmp309} \\ 
    &= \Pdl \sum_{\ell = 1}^L a_{\ell,i}  \frac{\eta_{k',\ell}}{N \gamma_{k',\ell}} c^2_{k',\ell} \Ppl K \Exop\left[\abs{\Delta_{k,\ell,i} \tp{\vecq}_{k,\ell}  \vecq_{k',\ell}}^2\right] \notag \\
    &\quad + \Pdl \sum_{\ell = 1}^L a_{\ell,i}  \frac{\eta_{k',\ell}}{N \gamma_{k',\ell}} c^2_{k',\ell}  \Exop\left[\abs{\Delta_{k,\ell,i} \tp{\vecq}_{k,\ell} \vecz\supp{pilot}_{k',\ell}}^2\right] \label{eq:tmp310} \\ 
    &= \Pdl \sum_{\ell = 1}^L a_{\ell,i}   \frac{\eta_{k',\ell}}{N \gamma_{k',\ell}} c^2_{k',\ell}  \Big(\Ppl K\Exop\left[\abs{\tp{\vech}_{k,\ell}  \vech_{k',\ell}}^2\right] \notag \\
    &\qquad + \Exop\left[\big|\tp{\vech}_{k,\ell} \vecz\supp{pilot}_{k',\ell}\big|^2\right] \Big)\label{eq:tmp337} \\
    &= \Pdl \sum_{\ell = 1}^L a_{\ell,i} \frac{\eta_{k',\ell}}{N \gamma_{k',\ell}} c^2_{k',\ell} N \beta_{k,\ell} \Big(\Ppl K  \beta_{k',\ell} + 1 \Big) \\
    &= \Pdl \sum_{\ell = 1}^L a_{\ell,i} \eta_{k',\ell} \beta_{k,\ell},
    \label{eq:UI}
\end{align}
where~\eqref{eq:tmp309} holds because $\vecz\supp{pilot}_{k',\ell}$ is independent of $\vecq_{k',\ell}$ and has zero mean, ~\eqref{eq:tmp310} holds because both $\Delta_{k,\ell,i} \tp{\vecq}_{k,\ell}  \vecq_{k',\ell}$ and $\Delta_{k,\ell,i} \tp{\vecq}_{k,\ell} \vecz_{k',\ell}$ are uncorrelated across $\ell$ and have zero mean. 

Finally, substituting~\eqref{eq:DS}, \eqref{eq:BU}, and~\eqref{eq:UI} into~\eqref{eq:rate_i}, we obtain the expression~\eqref{eq:rate_i_conj} of the achievable rate $R_{k,i}$.

\subsection{\gls{ZF} Beamforming}

The terms $|\DS_{k,i}|^2$ and $\Exop\left[\abs{\BU_{k,i}}^2\right]$ are given by the numerator and the first term in the denominator, respectively, of~\eqref{eq:rate_i_ZF}. We compute $\Exop\left[\abs{\UI_{k,i}}^2\right]$ as follows
\begin{align}
    \Exop\!\big[\abs{\UI_{k,i}}^2\big] 
    &= \Pdl \! \sum_{\ell = 1}^L  \! a_{\ell,i} \Exop\!\left[\abs{\Delta_{k,\ell,i} \tp{\tilde \vecq}_{k,\ell,i} \matW_{\ell,i} \matD_{\veceta_\ell}^{1/2} \vecs_{i} }^2 \right] \!\!\label{eq:tmp1470}\\
    &= \Pdl \sum_{\ell = 1}^L   a_{\ell,i}  (\beta_{k,\ell} - \gamma_{k,\ell}) \Exop\left[\vecnorm{ \matW_{\ell,i} \matD_{\veceta_\ell}^{1/2} \vecs_{i}}^2\right] \label{eq:tmp1554}
\end{align}
where \eqref{eq:tmp1470} follows because $\Delta_{k,\ell,i} \tp{\tilde \vecq}_{k,\ell,i} \matW_{\ell,i} \matD_{\veceta_\ell}^{1/2} \vecs_{i}$ are uncorrelated across $\ell$; in~\eqref{eq:tmp1554}, we used that $\tilde \vecq_{k,\ell,i}$ is distributed as $\jpg(\veczero, (\beta_{k,\ell} - \gamma_{k,\ell})\matidentity_N)$. We next have that
\begin{align}
    &\Exop\left[\vecnorm{ \matW_{\ell,i} \matD_{\veceta_\ell}^{1/2} \vecs_{i}}^2\right] \notag \\
    &= \Exop\left[\tr\big(\matW_{\ell,i} \matD_{\veceta_\ell} \herm{\matW}_{\ell,i}\big)\right] \\ 
    &= (N-K) \Exop\left[\tr\left(\matD_{\veceta_\ell} \big(\tp{(\widehat{\matQ}_{\ell,i} \matD_{\vecgamma_\ell}^{-1/2})} (\widehat \matQ_{\ell,i}\matD_{\vecgamma_\ell}^{-1/2}) ^*\big)^{-1}\right)\right] \notag \\
    &= \sum_{k'=1}^K \eta_{k',\ell}
\end{align}
where in the last equality, we used~\cite[Lemma~2.10]{Tulino04} to verify that the expectation term is equal to $(N-K)^{-1} \sum_{k' = 1}^K \eta_{k',\ell}$.
We conclude that 
\begin{equation}
    \Exop\left[\abs{\UI_{k,i}}^2\right] = \Pdl \sum_{\ell = 1}^L   a_{\ell,i} (\beta_{k,\ell} - \gamma_{k,\ell}) \sum_{k' = 1}^K \eta_{k',\ell}.
\end{equation} 
This completes the proof of~\eqref{eq:rate_i_ZF}.

\end{appendices}

\bibliographystyle{IEEEtran}
\bibliography{IEEEabrv,./ref}

\end{document}